\documentclass[twocolumn,floatfix, nofootinbib,prd,preprintnumbers,showpacs,showkeys,superscriptaddress,longbibliography]{revtex4-1}

\usepackage[mathscr]{euscript}
\usepackage{amsmath}
\usepackage{graphicx}
\usepackage{dcolumn}
\usepackage{bm}
\usepackage{epsfig}
\usepackage{amssymb,latexsym,mathrsfs}
\usepackage{graphicx}
\usepackage{color}
\usepackage{hyperref}
\usepackage{float}
\usepackage[thinlines]{easytable}
\usepackage{multirow}
\usepackage{diagbox}

\newcolumntype{M}[1]{>{\centering\arraybackslash}m{#1}}
\newcolumntype{N}{@{}m{0pt}@{}}

\usepackage{tikz}

\usepackage{amsmath}
\newcommand*\diff{\mathop{}\!\mathrm{d}}
\newcommand*\Diff[1]{\mathop{}\!\mathrm{d^#1}}

\hypersetup{
    colorlinks=true,
    linkcolor=red,
    citecolor=blue,
}

\usepackage{csquotes} 
\usepackage{amssymb}
\usepackage{physics}
\usepackage{subfigure}
\usepackage{hyperref}
\usepackage{url}
\usepackage{xcolor}
\usepackage{color}
\definecolor{amaranth}{rgb}{0.9, 0.17, 0.31}
\definecolor{purple(munsell)}{rgb}{0.62, 0.0, 0.77}
\definecolor{americanrose}{rgb}{1.0, 0.01, 0.24}
\definecolor{palatinateblue}{rgb}{0.15, 0.23, 0.89}
\definecolor{royalblue(web)}{rgb}{0.25, 0.41, 0.88}
\definecolor{hanpurple}{rgb}{0.32, 0.09, 0.98}
\definecolor{beaublue}{rgb}{0.74, 0.83, 0.9}
\definecolor{carminered}{rgb}{1.0, 0.0, 0.22}
\definecolor{brightpink}{rgb}{1.0, 0.0, 0.5}
\definecolor{vividviolet}{rgb}{0.62, 0.0, 1.0}

\definecolor{electron}{rgb}{1.0, 0.67, 0.22}
\hypersetup{ linktoc=all,
    colorlinks, linkcolor={palatinateblue},
    citecolor={brightpink}, urlcolor={amaranth}}

\newcommand{\be}{\begin{equation}}
\newcommand{\ee}{\end{equation}}
\newcommand{\bs}{\begin{split}} 
\newcommand{\bea}{\begin{eqnarray}}
\newcommand{\eea}{\end{eqnarray}}

\newcommand{\bes}{\begin{subequations}}
\newcommand{\ees}{\end{subequations}}

\renewcommand{\d}[1]{\ensuremath{\operatorname{d}\!{#1}}}

\renewcommand{\d}[1]{\ensuremath{\operatorname{d}\!{#1}}}

\newcommand{\bo}{\raise-1mm\hbox{\Large$\Box$}}

\newcommand{\bd}{\boldsymbol}

\begin{document}


\title{Larmor Temperature, Casimir Dynamics, and Planck's Law}

\author{Evgenii Ievlev}
\email{evgenii.ievlev@nu.edu.kz}
\altaffiliation[On leave of absence from: ]{National Research Center “Kurchatov Institute”, Petersburg Nuclear Physics
Institute, St.\;Petersburg 188300, Russia}
\affiliation{Physics Department \& Energetic Cosmos Laboratory,\\ Nazarbayev University,
Astana 010000, Qazaqstan}
\affiliation{Theoretical and Nuclear Physics Department, \\
al-Farabi Qazaq National University,
Almaty 050040, Qazaqstan}

\author{Michael R.R. Good}
\email{michael.good@nu.edu.kz}
\affiliation{Physics Department \& Energetic Cosmos Laboratory,\\ Nazarbayev University,
Astana 010000, Qazaqstan}
\affiliation{Leung Center for Cosmology and Particle Astrophysics,
National Taiwan University, Taipei 10617, Taiwan}

\begin{abstract} 
Classical radiation from a single relativistically accelerating electron is investigated where the temperature characterizing the system highlights the dependence on acceleration. In the context of the dynamic Casimir effect with Planck-distributed photons and thermal black hole evaporation, we demonstrate analytic consistency between the ideas of constant acceleration and equilibrium thermal radiation. For ultra-relativistic speeds, we demonstrate a long-lasting constant peel acceleration and constant power emission, which is consistent with the idea of balanced equilibrium of Planck-distributed particle radiation.

\end{abstract} 

\date{\today} 

\maketitle

\tableofcontents


\section{Introduction}
\label{sec:intro}
It is fascinating that black holes, with surface gravity $\kappa = c^4/4GM$, have `quantum' ($\hbar$) temperature \cite{Hawking:1974sw},
\be T_{\textrm{BH}} = \frac{\hbar \kappa }{2\pi c k_B},\label{bhtemp}\ee
because, in part, the radiated particles in equilibrium are frequency distributed with a Planck factor, and the power emitted scales according to $P\sim T^2$ substantiating black holes as one-dimensional information channels \cite{Bekenstein:2001tj}. 

In this note we help make the case for a classical analog to Eq.~(\ref{bhtemp}).  We present new details supporting the idea of a moving point charge radiation effect, quite similar in form to Eq.~(\ref{bhtemp}) yet fully classical in origin. The thermal radiation originates from a single accelerating electron.  For clarity, we provide overlap with \cite{Good:2022eub} but the novel results here focus on temperature and the analytic expressions of time-dependence. Our results are concerned with the equilibrium period of an electron's radiation: when the power emitted is uniform and classical thermodynamics applies, the emission has a temperature proportional to the peel acceleration, $\kappa$, of the electron; the latter term is defined and explained in Sec.~\ref{sec:peel}.  One finds the `classical' (no $\hbar$)\footnote{The temperature is in the Stoney scale \cite{stoney1881physical}; see Barrow \cite{barrowSTONE}.} temperature \cite{Ievlev:2023inj},
\be T_{\textrm{electron}} = \frac{\mu_0 e^2 \kappa}{2\pi k_B},\label{temp1}\ee
which is commensurate with constant power emission \cite{Good:2022eub}. Interestingly, this occurs during Planck-distributed radiation from an analog moving mirror (dynamical Casimir effect \cite{moore1970quantum,DeWitt:1975ys,Davies:1977yv}) accelerated along the same specific trajectory (given in \cite{Good:2022eub}).  A horizontal leveling of the power is visually seen at extremely ultra-relativistic final speeds of the electron.  A notion of temperature is congruent with the power emitted by the electron scaling according to $P\sim T^2$ revealing similar Bekenstein one-dimensional behavior and power-temperature scaling \cite{Bekenstein:2001tj}. In the following work, we investigate the arguments supporting this conjecture and the classical temperature Eq.~(\ref{temp1}), as well as the analogy to the quantum temperature Eq.~(\ref{bhtemp}). 

\subsection{Analog bridge}
\label{sec:analog_bridge}
First, let us consider the simple action-correspondence, between Eq.~(\ref{bhtemp}) and Eq.~(\ref{temp1}),
\be \hbar \to \frac{e^2}{\epsilon_0 c} = \mu_0 c e^2, \label{bridge} \ee
where the reduced Planck's constant is, as usual,
\be \hbar = 1.054 \times 10^{-34} \textrm{ J s},\ee
and the smaller action (or angular momentum) classical quantity is,
\be \mu_0 c e^2 = 9.671 \times 10^{-36} \textrm{ J s}.\ee
%
Notice that
\begin{equation}
    \frac{\hbar}{\mu_0 c e^2}  \approx 10.91.
\end{equation}
For a given acceleration scale $\kappa$, the classical temperature, Eq.~(\ref{temp1}) is nominally about a magnitude order smaller than the quantum temperature Eq.~(\ref{bhtemp}).  This analog `substitution', Eq.~(\ref{bridge}), if you will, can help us bridge the analog connection between the elementary particle and black hole in an easy-to-use way; i.e. substituting $\hbar \to \mu_0 c e^2$ in Eq.~(\ref{bhtemp}) gives Eq.~(\ref{temp1}).  We will help justify and generalize this in the following sections.

\subsection{Temperature definition}

Temperature is a collective property and is almost always defined with an assemblage of particles. The usefulness of thermodynamics is particularly salient in the regime with a large numbers of particles (in this case, the large amount of radiated particles are infinite soft thermal photons).

We emphasize that what is meant by `the temperature of  electron radiation' in Eq.~\eqref{temp1} is a temperature extracted by averaging the photon energy radiated over many realizations of the
same decay experiment with a single asymptotically ultra-relativistic electron.  Only in this context, does it makes sense to consider a single electron radiating photons with a defined temperature. 

Here the frequency-distribution is also analogous to the moving mirror particle production which is Planck-distributed, see Sec.~\ref{sec:planck} below.  The connection to black hole temperature is limited in the sense that an explicit Planck-distribution has not been derived for classical electron radiation, unlike the moving mirror Planck-distribution which is a result of the beta Bogolubov coefficients originating from the quantum fields in curved space approach. However the connection is explicitly tethered by the power-temperature scaling of the (1+1) dimensional Stefan-Boltzmann law, see Sec.~\ref{sec:sb} below.  Importantly, this notion of electron radiation temperature is dynamically useful because it signals a corresponding period of uniform peel acceleration (which we define in Sec.~\ref{sec:peel}).

\subsection{Extension bridge}
The bridge, Eq.~(\ref{bridge}), is not limited to Eq.~(\ref{bhtemp}) and Eq.~(\ref{temp1}).  It proves useful as an action correspondence in general (c.f. Ritus~\cite{Ritus:2003wu,Ritus:2002rq,Ritus:1999eu,Nikishov:1995qs}) between the quantum moving mirror model ($q$) and the classical moving point charge model ($c$). This is seen, respectively, in the power, see e.g. \cite{Zhakenuly:2021pfm} and \cite{Good:2021ffo}, where $\alpha$ is the proper acceleration of the mirror or electron,
\be P_q = \frac{\hbar \alpha^2}{6\pi c^2}, \quad P_c = \frac{\mu_0 e^2 \alpha^2}{6\pi c},\ee
and self-force \cite{Ford:1982ct,Unruh:1982ic}, where the prime indicates the derivative with respect to proper time, $\tau$, (not coordinate time $t$) see e.g. \cite{Myrzakul:2021bgj} and \cite{Ford:1982ct},
\be F_q = \frac{\hbar \alpha'(\tau)}{6\pi c^2}, \quad F_c = \frac{\mu_0 e^2 \alpha'(\tau)}{6\pi c},\ee
for any 
limited, horizon-less trajectory whose acceleration is asymptotically zero (asymptotic inertia). 

Moreover, the bridge also occurs specifically between the spectral radiance of a particular moving mirror model and lowest order inner bremsstrahlung (IB) during beta decay \cite{Good:2022eub}, contained therein. More generally, if we examine the infrared limit for example, this is seen, respectfully, in the frequency independence of the spectral energy per unit bandwidth, see e.g. \cite{Zangwill:1507229} and \cite{Zhakenuly:2021pfm},
\be I_q = \frac{\hbar}{2\pi^2}\left(\frac{\eta}{s} - 1\right), \quad I_c = \frac{\mu_0 c e^2}{2\pi^2}\left(\frac{\eta}{s} - 1\right),\label{spectra1}\ee
where $s = \tanh\eta$ is the final speed of the mirror or electron as a fraction of the speed of light $c$, and $\eta$ is the final rapidity.

Below we are going to check the consistency of our claims by analyzing this correspondence from different sides.
In Sec.~\ref{sec:electron_radiates} we consider the radiation from an accelerated electron more closely, discussing the relevant scales of the problem.
In Sec.~\ref{sec:plateau} we will show that such an electron is characterized by constant-in-time characteristic quantities, thus supporting the thermal regime.
Sec.~\ref{sec:planck} presents the  Planck spectrum for a moving mirror model and its connection to the moving point charge and black holes, also supporting thermality and the analog bridge of Eq.~(\ref{bhtemp}) and Eq.~(\ref{temp1}).
in Sec.~\ref{sec:sb} we provide several derivations of the Stephan-Boltzmann law in the relevant contexts; the results serve as an independent confirmation of the quadratic dependence in temperature.
Sec.~\ref{sec:concl} presents our conclusions.

\section{Energy radiated by an electron}
\label{sec:electron_radiates}

\subsection{Total energy emitted}
To obtain the energy per unit bandwidth from Eq.~(\ref{spectra1}), one associates the UV scale of the system with the acceleration scale $\kappa$,
\begin{equation}
    \omega_{\textrm{max}} = \frac{\pi \kappa}{12c}, \label{omegamax_kappa_relation}
\end{equation}
such that, using the first equation of Eq.~(\ref{spectra1}), the quantum spectral energy per unit bandwidth,
\be I_q = \frac{\diff E_q}{\diff \omega} \quad \to \quad E_q = \frac{\hbar \kappa}{24\pi c}\left(\frac{\eta}{s} - 1\right).\ee
or the second equation of Eq.~(\ref{spectra1}), the classical spectral energy per bandwidth,
\be I_c = \frac{\diff E_c}{\diff \omega} \quad \to \quad E_c = \frac{\mu_0 e^2 \kappa}{24\pi }\left(\frac{\eta}{s} - 1\right).\label{classicalenergy}\ee
With the clarity of SI units, this demonstrates that the two different models have analogous energy emission scaling.  Notice that the energy Eq.~(\ref{classicalenergy}) can be expressed as 
\be E_c =  \frac{\mu_0 e^2 \kappa}{48\pi} \left[\frac{1}{s}\ln \left(\frac{1+s}{1-s}\right)-2\right],\label{EnergyElectron1} \ee
where again, $s$ is the final constant speed of the electron as a fraction of the speed of light, and $\eta$ is the rapidity from $s = \tanh \eta$.  As we shall see, for a thermal plateau, ultra-relativistic speeds are required, $s\sim 1$ (although this is not required in order to obtain the Planck distributed photon thermal spectrum). Only the classical version Eq.~(\ref{classicalenergy}) \cite{Jackson:490457} or lowest order IB energy \cite{PhysRev.76.365}, has been directly observed in experiments, see e.g. \cite{Ballagh:1983zr}.

\subsection{UV cut-off and temperature}
For some orientation, consider now, re-expressing the temperature of the electron radiation, Eq.~(\ref{temp1}), in terms of the maximum appreciable energy emitted, $\hbar \omega_{\textrm{max}} = \hbar \pi \kappa / 12 c$ (cf. Eq.~\eqref{omegamax_kappa_relation})
\begin{equation}
    T = \frac{6}{\pi^2} \frac{ \mu_0 c e^2 }{ \hbar  }   \frac{E_\gamma}{k_B} \,, \label{temp}
\end{equation}
where energy range of the detected photons is UV limited by $E_\gamma = \hbar \omega_{\textrm{max}}$ and can be expressed from Eq.~\eqref{temp} as,
\begin{equation}
    E_\gamma =  \frac{\pi^2}{6} \frac{ \hbar  }{ \mu_0 c e^2 } k_B T \approx 18\, k_B T.
\end{equation}
%
This gives some perspective of the dependence of the system on the cut-off when in thermal equilibrium. 
Consistency of the temperature formulas Eq.~\eqref{temp1} and Eq.~\eqref{temp} is confirmed in the following Sections.

Including these UV limits, experimental evidence of lowest order IB energy emitted during beta decay confirms the consistency of the theoretically derived frequency independence of the spectral energy per unit bandwidth, Eq.~(\ref{spectra1}), see e.g. \cite{Ballagh:1983zr}.  In the following sections, we support the physical notion of temperature in this context by providing corroborative analytic results confirming the mathematical validity of Eq.~(\ref{temp1}).




\subsection{Scale dependence}
The analog between black hole temperature and electron radiation temperature, introduced in Sec.~\ref{sec:intro}, has limitations.  Black hole temperature, $T=\hbar \kappa/2\pi c k_B$, varies dependent on the surface gravity, $\kappa = c^4/4GM$, of the black hole, while electron radiation temperature, $T = \mu_0 e^2 \kappa/2\pi k_B$, varies on the acceleration scale, 
$\kappa = 12 c\omega_\text{max}/\pi$
inherently a function of the 
UV scale of the system, $\omega_\text{max}$.
Hence, because the charge of every electron is the same, the fine structure does not change in this context, and the temperature of the electron's acceleration radiation is 
UV
dependent, the two expression differ with respect to both intuition and scale.  In this context, it is useful to consider the universality of the soft-factor \cite{Strominger:2017zoo} and the thermal character of the infinite zero-energy photons emitted in this regime.  Indeed, the thermality here is connected to every scattering process in the deep infrared, at least in the instantaneous collision reference frame \cite{Cardoso:2003cn}. Thus there is an argument for the relevance of Eq.~(\ref{temp1}) beyond the bremsstrahlung context. 

To this end, we point out that Eq.~(\ref{temp1}) is relevant for  Feddeev-Kulish dressed states, where equivalent particle count and energy results \cite{Tomaras} suggest one can can derive a `cloud temperature'.  Analog systems with corresponding results are also subject to thermal character.  For instance, `mirror temperature' is a useful assignment in the context of the dynamical Casimir effect \cite{Good:2016yht}, as we have directly demonstrated with the spectral computation Eq~(\ref{betaT1}). Moreover, since the internal structure of the source cannot be discerned by long wavelengths, these results can necessarily be extended in analog to curved spacetime final states \cite{wilczek1993quantum} where `black hole temperature' leading to a left-over remnant becomes a useful characterization of the system.  We leave these extensions for future investigations.

\section{Thermal plateaus}
\label{sec:plateau}

In a system with well-defined thermality one naturally expects to see an equilibrium, which implies that characteristic quantities describing this system remain constant in time (up to small fluctuations).
For example, a black body immersed in a heat bath radiates the same amount of energy each second, i.e. the radiation power remains constant.

Since we've talked about temperature and thermality for the electron/mirror setup, it is desirable to see that this system is indeed in a regime where its characteristic quantities remain constant.
In this section we explore this in detail.
We find that indeed there is a time window where the characteristic quantities remain constant and exhibit plateaus.
In the ultrarelativistic limit of high final speeds $s \to 1$ these plateaus become very wide, thus validating the regime of thermality.

\subsection{Constant power emission}
As expected for thermal equilibrium, a stable emission period of constant power is measured by a far-away observer.  This is best represented as the change of energy with respect to retarded time $u=t-r/c$, and written as Larmor power $\bar{P} = \frac{\diff E}{\diff u},$ such that
$\bar{P} = P \frac{\diff t}{\diff u} = P/(1-\beta)$, where $P = \mu_0 e^2 \alpha^2/6\pi c$. Here $\alpha$ is the proper acceleration and $\beta$ is the velocity normalized by the speed of light $c$.  

\begin{figure}[H]
\centering
  \centering
  \includegraphics[width=0.9\linewidth]{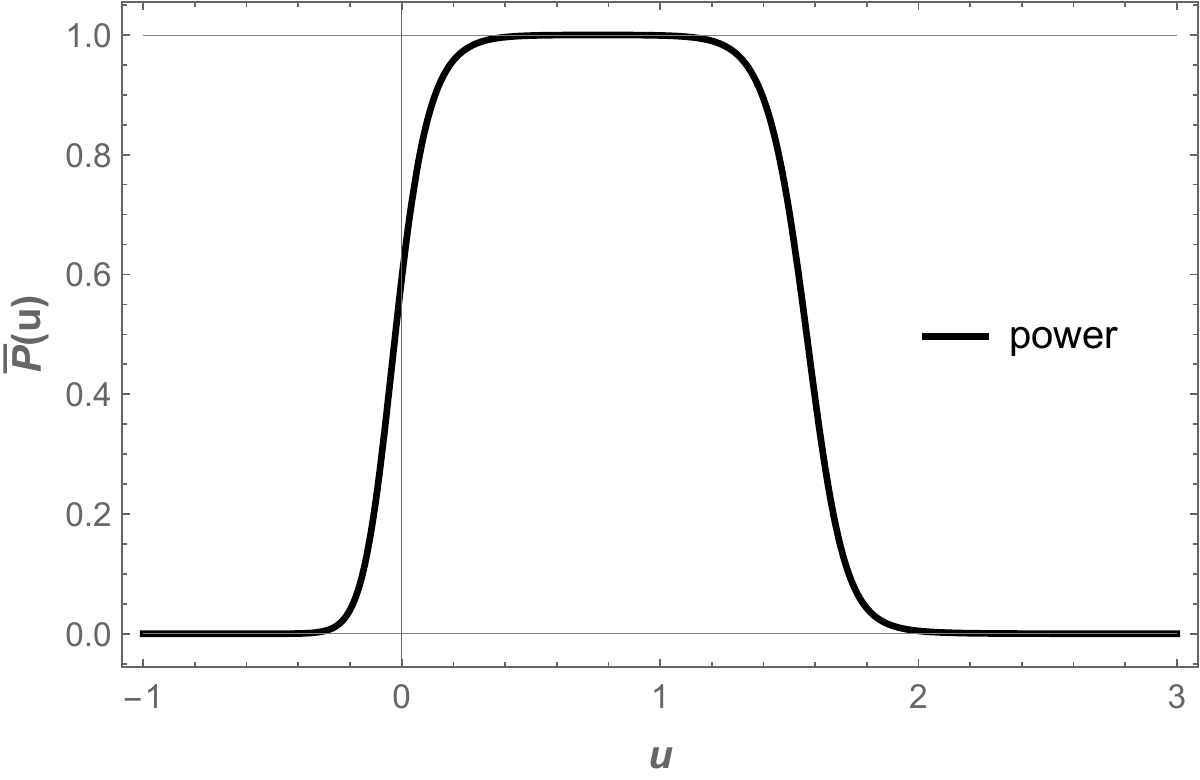}
 \caption{A plot of the power, $\bar{P}(u)$, Eq.~(\ref{barP}), with a plateau demonstrating constant emission when the final speed of the electron is extremely ultra-relativistic, $s=1-10^{-9}$ or rapidity $\eta = 10.7$. Here $\kappa =\sqrt{48\pi}$ and $\mu_0 = c = 1$ and unit charge so that the plateau is at height $\bar{P}(u) =1$. The Larmor power plateau corroborates the conclusion that at high electron speeds the photons, like the Planck-distributed particles Eq.~(\ref{betaT1}) produced by the mirror, find themselves with temperature $T=\kappa/2\pi$, Eq.~(\ref{temp1}).  The integral under the curve, Eq.~(\ref{barP}), is the experimentally observed soft IB energy, Eq.~(\ref{classicalenergy}) or Eq.~(\ref{EnergyElectron1}).   }
\label{power}
\end{figure}

Our main example will be the trajectory directly related to the lowest order inner bremsstrahlung in the radiative beta decay \cite{Good:2022eub},
\begin{equation}
    \bd{r}(t) = \frac{sc}{\kappa}W(e^{\kappa t/c})\bd{\hat{r}} \,,
\label{trajectory_r_of_t}
\end{equation}
Here, $W$ is the Lambert product logarithm defined as a solution to equation $w e^w = x$ such that $w = W(x)$ and $W(0)=0$.
The Larmor power can be computed analytically; it's expression $\bar{P}(u)$, formulated in terms of retarded time $u$, is 
\be \bar{P} = \frac{\mu_0 e^2\kappa ^2 s^2 W^2 (W+1-s)}{6 \pi c  (W+1)^4 ((1+s) W+1-s)^3},\label{barP}\ee
where $W= W[e^{\kappa  u/c}(1-s)]$ is again the Lambert product log.  This result has a plateau when the final speed of the electron is near the causal limit $s \to 1$.  Consider analytically, two separate limits, of high speeds and late times, which reveals, using Eq.~(\ref{temp1}),
\be \bar{P}_c \equiv \lim_{u\to\infty}\lim_{s\to 1} \bar{P}(u) = \frac{\mu_0 e^2\kappa^2}{48\pi c} = \frac{\pi}{12}\frac{k_B^2 }{\mu_0 c e^2}T^2.\label{SBlaw}\ee
A $\bar{P}(u)$ plot at high final asymptotic speeds $s \sim 1$ illustrates the constant power plateau indicative of thermal emission, Figure \ref{power}.

We keep in mind, that we are working with classical (3+1) dimensional radiation of an electron.  Therefore, we notice that Eq.~(\ref{SBlaw}) is a (1+1) dimensional classical power-temperature relation,
with scaling identical to the standard quantum (1+1) dimensional Stefan-Boltzmann law \cite{Landsberg1989TheSC} which describes (3+1) dimensional black hole power radiance, see e.g. \cite{Bekenstein:2001tj},
\be P_q = \frac{\pi k_B^2}{12\hbar}T^2.\label{SIPq}\ee
In the same way that a single spatial dimensional Planck distribution yields Eq.~(\ref{SIPq}), an analog Planck distribution $J$, absent $\hbar$, or spectral energy density in angular frequency space, where Eq.~(\ref{bridge}), $\hbar \to \mu_0 c e^2$, can be applied (as an example),
\be J(\omega) = \frac{1}{2\pi} \frac{\mu_0 ce^2 \omega}{e^{\mu_0 c e^2 \omega/k_B T}-1},\label{J}\ee
integrated over angular frequency,
\be \int_0^\infty J(\omega) \diff \omega = \frac{\pi}{12}\frac{k_B^2 }{\mu_0 c e^2}T^2,\ee
which results in Eq.~(\ref{SBlaw}).  It is natural to suppose a distribution similar to $J(\omega)$, Eq.~(\ref{J}), might be responsible for Eq.~(\ref{SBlaw}), see e.g. \cite{Ievlev:2023inj}.  Such a distribution could lend support for the action correspondence, Eq.~(\ref{bridge}), but also corroborate the temperature Eq.~(\ref{temp1}).  It appears that such a distribution would only characterize the radiation during a long-lived constant power emission phase at sufficiently high speeds $s\sim 1$. Nevertheless, independent of any $J(\omega)$ supposition and the difficulties commensurate with such speculation, the power emission, Eq~(\ref{barP}), possesses a plateau consistent with Eq.~(\ref{SBlaw}). 
\begin{figure}[H]
\centering
  \centering
  \includegraphics[width=0.9\linewidth]{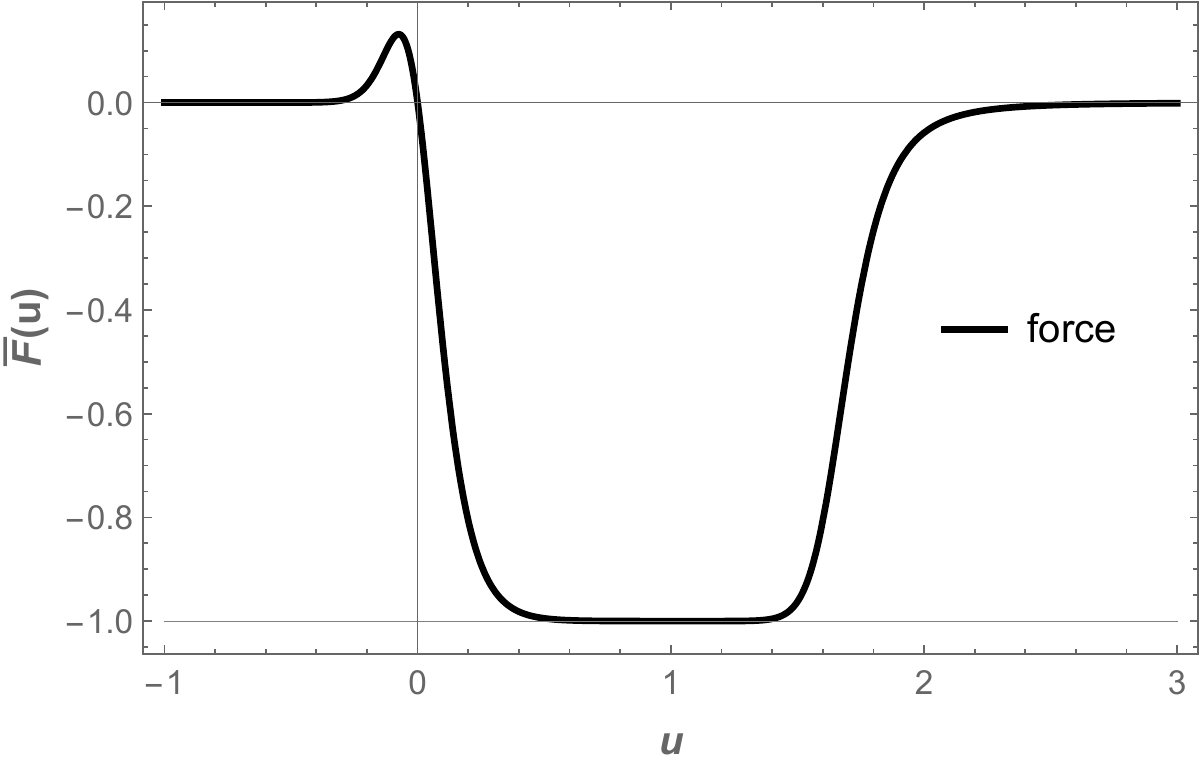}
 \caption{A plot of the Feynman power or `self-force measure', $\bar{F}(u)$, Eq.~(\ref{barF}), with a plateau demonstrating constant self-force when the final speed of the electron is extremely ultra-relativistic, $s=1-10^{-9}$ or rapidity $\eta = 10.7$. Here $\kappa =\sqrt{48\pi}$ and $\mu_0 = c = 1$ and unit charge so that the plateau is at height $\bar{F}(u) =-1$. The Feynman power plateau corroborates the  temperature $T=\kappa/2\pi$, Eq.~(\ref{temp}).   The integral under the curve, Eq.~(\ref{barF}), is sign-flipped  experimentally observed soft IB energy, Eq.~(\ref{EnergyElectron1}).  }
\label{force}
\end{figure}

\subsection{Constant radiation reaction}
Having seen the power plateau in $\bar{P}(u)$ originating from $P=\mu_0 e^2 \alpha^2/6\pi c$, we now turn to the self-force, $F=\mu_0 e^2 \alpha'(\tau)/6\pi c$ (where $\tau$ is proper time) and the associated power which we call `Feynman power' \cite{Feynman:1996kb}, $\bar{F}(u) = F\frac{\diff r}{\diff u} = F \beta/(1-\beta)$, as a function of retarded time $u$,
\be \bar{F} = \frac{\mu_0 e^2 \kappa ^2 s W (s-W-1) \left(2 (s+1) W^2+s+W-1\right)}{6 \pi c  (W+1)^4 ((s+1) W-s+1)^3}. \label{barF}\ee
We have again used $W= W[e^{\kappa  u/c}(1-s)]$, advanced coordinate $v= t + r/c$, and retarded time coordinate $u = t-r/c$.  Taking the same two separate consecutive limits of high speeds and late times, as done for Larmor power in Eq.~(\ref{SBlaw}), reveals,
\be \lim_{u\to\infty}\lim_{s\to 1} \bar{F}(u) = -\frac{\mu_0 e^2 \kappa^2}{48\pi c} = -\frac{\pi}{12} \frac{k_B^2}{\mu_0 c e^2} T^2.\label{twolimits}\ee
See a plot of the period of constant Feynman power in Figure \ref{force}. It, like the Larmor power $\bar{P}$, also exhibits a constant period during which the electron emits particles in thermal equilibrium.   Eq.~(\ref{twolimits}) substantiates Eq.~(\ref{temp1}).

\subsection{Constant peel acceleration}
\label{sec:peel}
Direct corroboration of an extended period of thermal equilibrium is given by the object: $\bar{\kappa}(u)= \partial_u \ln v'(u)$.  
This quantity is called the `peeling function' and has been used in the relativity literature, see e.g. \cite{Bianchi:2014qua,Barcelo:2010pj}. 
Following precedent, we call it the peel acceleration or `peel' for short.

The peel acceleration typically accompanies thermal particle radiation.  For instance, it has been used as a measure of what Carlitz-Willey \cite{CW2lifetime} called `local acceleration'.  The result for IB is \cite{Good:2022eub},
\be \bar{\kappa}(u) = \frac{2 \kappa  s W}{(W+1)^2 (1+(s+1)W-s)}, \label{locala}\ee
where $W= W[e^{\kappa  u/c}(1-s)]$. In the limit of high speeds and late times one sees,
\be \lim_{u\to\infty} \lim_{s\to 1} \bar{\kappa}(u) = \kappa.\ee
The peel acceleration, $\bar{\kappa}(u)$, is related to the Lorentz invariant proper acceleration, $\alpha$ via the relations $\bar{\kappa} = 2\alpha e^\eta$ or via the first derivative of the rapidity with respect to retarded time, $\bar{\kappa}(u) = 2\eta'(u)$. 

\begin{figure}[H]
\centering
  \centering
  \includegraphics[width=0.9\linewidth]{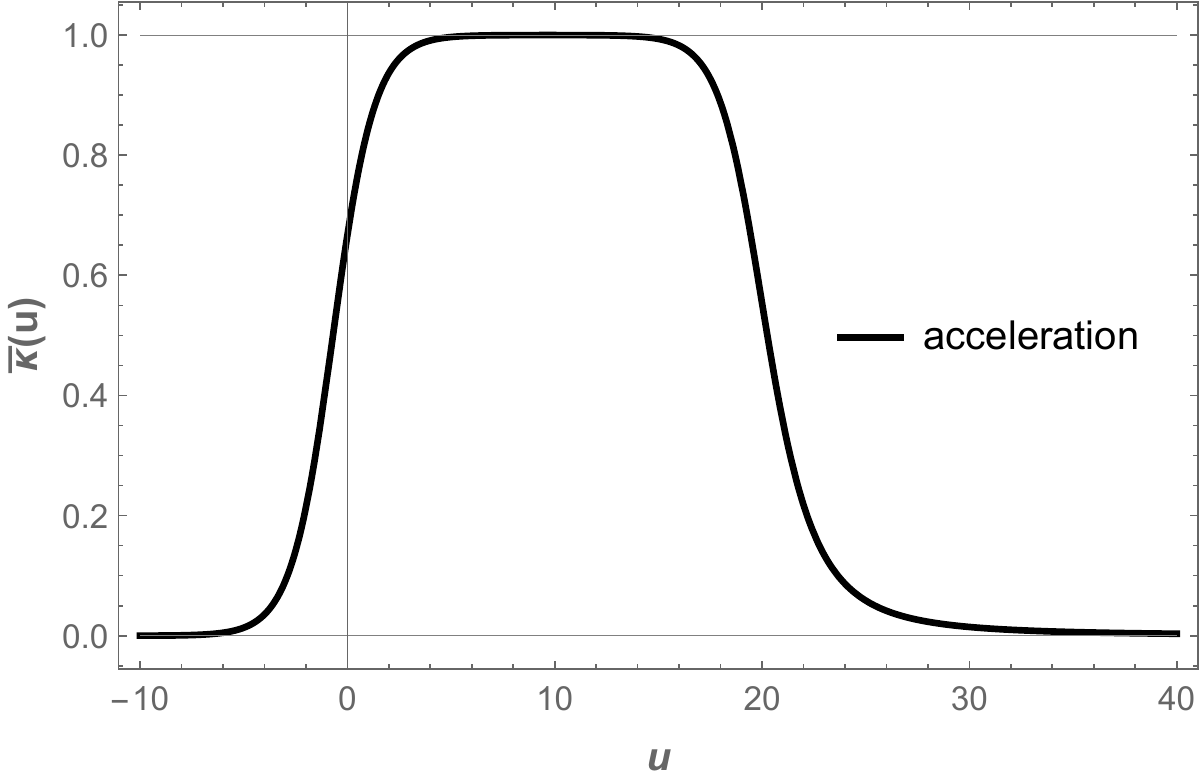}
 \caption{A plot of the peel acceleration, Eq.~(\ref{locala}), $\bar{\kappa}(u)$, with a plateau demonstrating constant local acceleration when the final speed of the electron is extremely ultra-relativistic, $s=1-10^{-9}$ or rapidity $\eta = 10.7$. Here $\kappa = c = 1$  so that the plateau is at height $\bar{\kappa}(u) =1$. The peel acceleration plateau directly substantiates the temperature $T=\kappa/2\pi$, Eq.~(\ref{temp1}).    }
\label{local}
\end{figure}

A plot of the peel acceleration is given in Figure \ref{local}.  A quasi-constant peel acceleration is in harmony with the equilibrium of a thermal distribution and constant power emission; however, it is important to underscore the fact that a constant peel acceleration does not describe uniform proper acceleration of the electron.


\section{Planck spectrum}
\label{sec:planck}

\subsection{Moving mirror model}

\begin{figure}
\centering
  \centering
  \includegraphics[width=0.8\linewidth]{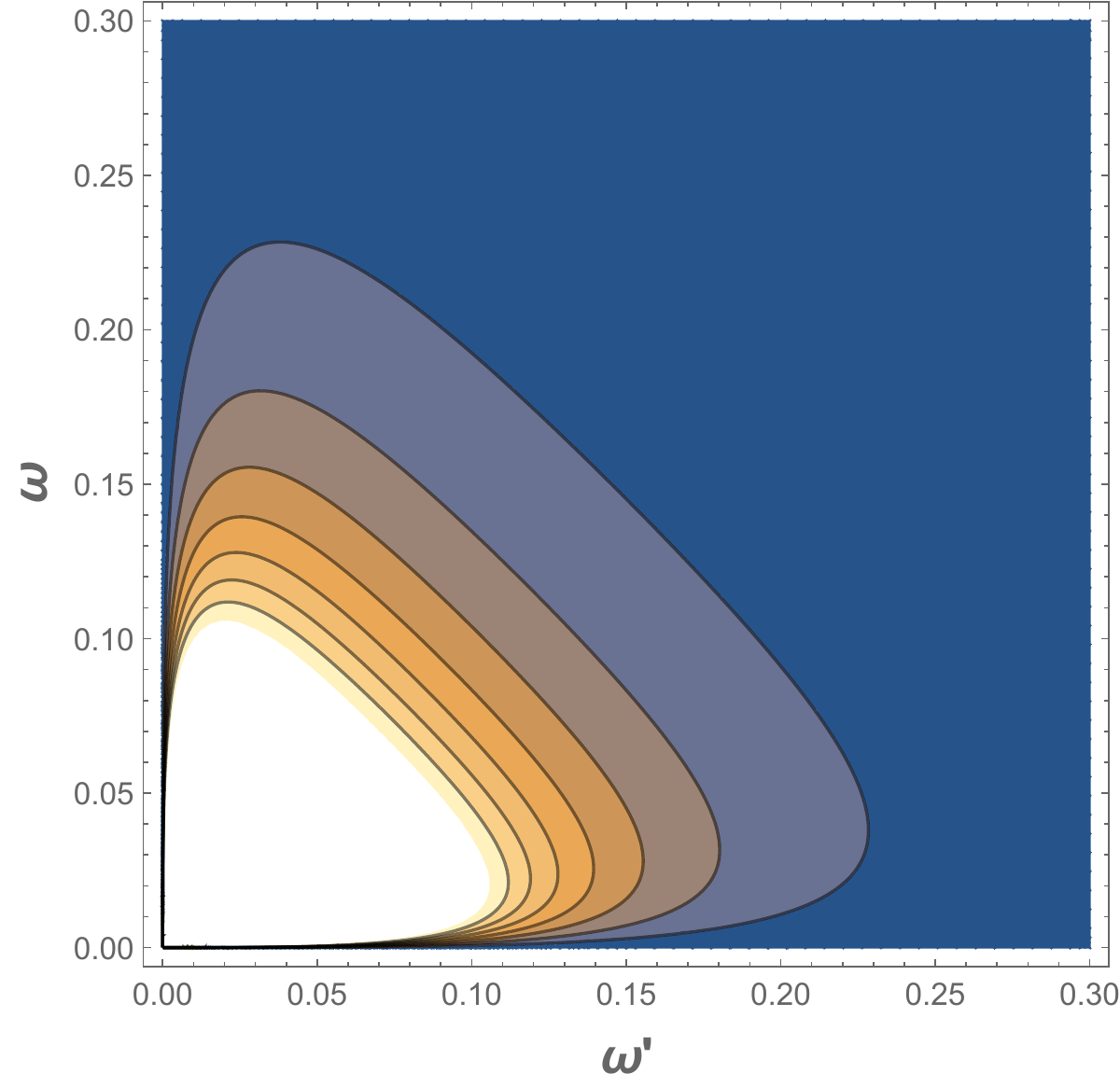}
 \caption{The $|\beta_{\omega\omega'}|^2$ spectrum of Eq.~(\ref{betaT1}) in a contour plot. Here $\kappa=1$ and $s=1/2$. Notice the symmetry between the frequency modes $\omega$ and $\omega'$. The qualitative shape is indicative of the Bose-Einstein statistics explicit in the Planck factor of Eq.~(\ref{betaT1}).}
\label{contour}
\end{figure}

In the moving mirror model (see e.g. \cite{Davies:1976hi,Davies:1977yv}), the beta Bogolubov coefficients corroborate radiative equilibrium via an explicit Planck distribution.  For IB during beta decay the Planck distribution is explicitly manifest in Eq.~(\ref{betaT1}).  Accelerating boundaries radiate soft particles whose long wavelengths lack the capability to probe the internal structure of the source \cite{Strominger:2017zoo}.  In the spirit of analogy, the moving mirror spectrum, with peel acceleration $\kappa$, is a good way to support the useful notion of temperature for the soft-spectrum of the electron's IB.  Combining the results for each side of the mirror \cite{Good:2016yht} by adding the squares of the beta Bogolubov coefficients, the overall spectrum is \cite{Good:2022eub}
\be |\beta_{\omega\omega'}|^2 = \frac{2 c s^2 \omega  \omega '}{\pi  \kappa  (\omega+\omega') }\frac{ a^{-2} + b^{-2}}{ e^{2\pi c (\omega+\omega')/\kappa}-1}.\label{betaT1}\ee
Here $a= \omega(1+s) + \omega'(1-s)$, and $b=\omega(1-s)+\omega'(1+s)$. See Figure \ref{contour} for an illustration of the symmetry between the modes.

For a consistency check, using the retarded time clock of the observer, the following integrations hold:
\be E_c= -\int_{-\infty}^{\infty} \bar{F}(u) \diff u = \int_{-\infty}^{\infty} \bar{P}(u) \diff u, \label{barPF}\ee
along with 
\be E_q = \int_0^\infty \int_0^\infty \hbar \omega |\beta_{\omega\omega'}|^2 \d \omega \d \omega', \label{numerical}\ee
or
\be E_q = \int_0^\infty \int_0^\infty \hbar \omega' |\beta_{\omega\omega'}|^2 \d \omega \d \omega', \label{numerical_prime}\ee
demonstrating consistency with the conservation of energy.  Importantly, this also demonstrates the consistency of the analogy between quantum mirrors and classical electrons.

The Planck spectrum of Eq.~(\ref{betaT1}) is robust to both high frequency and low frequency approximations.  This is made particularly explicit by considering high final speeds, $s\approx 1$, then using either frequency approximation.  Consider the high frequency, $\omega'\gg \omega$ approximation \cite{Hawking:1974sw}.  To leading order, one retrieves
\begin{equation}
    |\beta_{\omega\omega'}|^2 = \frac{c}{2 \pi  \kappa  \omega} \frac{1}{e^{2\pi c \omega'/\kappa} -1}
\label{mirror_planck_spec_1}
\end{equation}
Likewise, considering high final speeds and the low frequency approximation, $\omega' \ll \omega$ switches the prime on the $\omega$'s, leading to the (see e.g. \cite{Fulling_optics})
\be 
|\beta_{\omega\omega'}|^2 = \frac{c}{2 \pi  \kappa  \omega'} \frac{1}{e^{2\pi c \omega/\kappa} -1}, \ee
demonstrating Planck-factor validity to either frequency approximation.  
%
The spectrum plot of the moving mirror radiation (Figure \ref{contour}) illustrates the explicit Planck factor which demonstrates the particles, $N(\omega) = \int d\omega'|\beta_{\omega\omega'}|^2$, are distributed with a temperature given in Eq.~(\ref{temp1}).

\subsection{Relation to electrons and black holes}

The moving mirror model, or dynamical Casimir effect (DCE), is closely related to electron radiation and black holes. 
Having the radiation spectrum of the mirror, it is possible to obtain the radiation spectra for these related systems.
Let us explain the details.

The connection between DCE and point charge radiation has been suggested long ago (see Unruh-Wald \cite{Unruh:1982ic} and Ford-Vilenkin \cite{Ford:1982ct}), 
and has been developing since, see \cite{Nikishov:1995qs,Ritus:1999eu,Ritus:2002rq,Ritus:2003wu,Zhakenuly:2021pfm} and \cite{Ritus:2022bph}.
Eventually, this led to the realization that there is an exact functional identity between the radiation spectra in these models \cite{Ievlev:2023inj,Ievlev:2023bzk}. 
In the latter papers, the corresponding transformation recipe was derived and checked; it was established that an electron corresponding to the mirror Eq.~\eqref{betaT1} radiates with the spectrum
\begin{equation}
	I(\omega) = \frac{\mu_0 c e^2}{2\pi^2}\left(\frac{ \tanh^{-1}s/c}{s/c}-1\right) \frac{2\pi c \omega/\kappa}{e^{2\pi c \omega/\kappa}-1}.
\label{I_result_1}
\end{equation}
One can immediately see the aforementioned Planck form of the spectrum with the same temperature as the mirror.

Thermal emission is not so surprising considering the Larmor power plateau (Figure \ref{power}), Feynman power plateau  (Figure \ref{force}), and acceleration plateau (Figure \ref{local}).  It is also in harmony with the close analogy for quantum and classical quantities of powers  \cite{Good:2021ffo,Zhakenuly:2021pfm} and self-forces \cite{Myrzakul:2021bgj,Ford:1982ct} between mirrors and electrons. 


Black hole evaporation \cite{Hawking:1974sw} and in particular, the collapse of a null shell in the $s$-wave approximation can also be described as a DCE \cite{wilczek1993quantum,Cong:2018vqx}.
This black hole -- moving mirror correspondence has been successfully applied, for example, to such important spacetimes as Schwarzschild \cite{Good:2016oey,Good:2018zmx}, Reissner–Nordstr\"om \cite{good2020particle}, and Kerr \cite{Good:2020fjz}.
In the triarchy `moving mirrors -- electrons -- black holes', the quantum-classical temperature relation between Eq.~(\ref{bhtemp}) and Eq.~(\ref{temp1}) is found, supporting the analog bridge of Sec.~\ref{sec:analog_bridge}.  It is an interesting question about which geometry corresponds to the mirror Eq.~\eqref{betaT1}; we leave this for a future investigation.


\section{Stefan-Boltzmann law}
\label{sec:sb}
It is natural to consider how the classical power scales according to the (1+1) dimensional Stefan-Boltzmann law  \cite{Landsberg1989TheSC},
\be P \sim T^2,\ee
rather than the (3+1) Stefan-Boltzmann law, 
\be P \sim A T^4,\ee
which governs\footnote{In flat spacetime, this is the relevant contrasting expression for the energy transmission of a single photon polarization out of a closed hot black body surface with temperature $T$ and area $A$ into 3–D space.} the power radiated from a black body in terms of its temperature. A first heuristic answer is the classical electron is a point particle with no area.

Ultimately, a better understanding may be related to black hole radiance.  
The scaling could occur for the same reason black holes are one-dimensional information channels \cite{Bekenstein:2001tj}, whose power also scales according to $P\sim T^2$.  
In the context of Eq.~(\ref{temp1}), the electron's constant power peaks at exactly Eq.~(\ref{SBlaw}) which is the analog of the well-known all-time constant equilibrium emission of the quantum stress tensor for the eternal thermal Carlitz-Willey moving mirror \cite{carlitz1987reflections} and the late-time Schwarzschild mirror \cite{Good:2016oey}. The 1+1 spacetimes corresponding to these mirrors exhibit horizons and have been considered as analogous to black holes, see e.g. \cite{wilczek1993quantum}.

A complete investigation concerning the entropy and information flow related to the quadratic temperature dependence of the electron's power emission is a worthwhile study but is outside the scope of this work. Nevertheless, in the following subsections, we will make some necessary preliminary progress into exploring this Stefan-Boltzmann law in the context of its origin from electromagnetic spectral analysis, statistically maximized entropy, and classical thermodynamics.

\subsection{Classical Stefan-Boltzmann}
Using the aforementioned Stoney scale \cite{stoney1881physical}, the classical temperature of radiation from an electron is, Eq.~(\ref{temp1}),
\be T = \frac{\mu_0 e^2 \kappa}{2\pi k_B}.\label{Temp}\ee
Contrast this with the Kelvin scale and the temperature resembles the quantum Davies-Fulling-Unruh effect,
\be T= \frac{\hbar \kappa}{2\pi c k_B},\ee
except here $\kappa$ is the peel (not uniform proper acceleration). The Davies-Fulling-Unruh expression is well-understood as a quantum effect and the proposed temperature of radiation emitted by an electron in the literature, e.g. \cite{Bell:1982qr,Myhrvold:1983hv,Kolbenstvedt_2001,Landulfo:2019tqj}. 

However, the classical reasoning for Eq.~(\ref{Temp}) is two-fold: dynamics and spectral analysis. Dynamically one can compute the power \cite{Good:2022eub} and find it agrees with the Stefan-Boltzmann law, $P \sim T^2$, at the plateau for high speeds, $s\approx 1$.  The spectral analysis below will confirm the trans-Planckian distribution; found in \cite{Good:2022eub}, using the spectrum $ I(\omega) = \diff{E}/\diff{\omega}\diff{\Omega}$ \cite{Ievlev:2023inj},
\be I(\omega) = \frac{\mu_0 c e^2}{2\pi^2}\left(\frac{\eta}{s} - 1\right) \frac{M}{e^M - 1},\ee
where the dimensionless $M$ is an analog to $\hbar \omega/(k_B T)$:
\be M \equiv \frac{\mu_0 c e^2}{k_B T} \omega, \ee
where the temperature is given by Eq.~(\ref{Temp}). Moreover, as we shall see, the characteristic frequency of the photons confirms the Stefan-Boltzmann law using basic classical electromagnetic spectral analysis.

\subsection{Stefan-Boltzmann from Spectra}
Allow us to assume thermal emission is described by a heuristic and characteristic frequency of the radiation when the electron is ultra-relativistic.  This frequency will be 
\be \frac{P}{k_B T} = \int_0^\infty \mathcal{I}(f) \diff f,\label{charfreq}\ee
where the left-hand side is the ratio of the thermal power divided by the average energy in equilibrium, $k_B T$, as given by the equipartition theorem for the canonical ensemble.  Here, 
\be \mathcal{I}(f) = \frac{I(f)}{I_\textrm{infra}} = \frac{M}{e^M - 1},\ee
is the dimensionless spectrum, as a function of frequency $f = \omega/(2\pi)$, so that $M = \frac{\mu_0 c e^2}{k_B T} 2\pi f$.  Here $I_\textrm{infra}$ is the infrared limit of the spectrum; see Eq.~(58) in \cite{Ievlev:2023inj} and Eq.~\eqref{spectra1} here.
Integrating Eq.~(\ref{charfreq}) over $f$ gives the required result for the power, 
\be P = \frac{\pi}{12} \frac{k^2_B}{\mu_0 c e^2}{T^2},\label{SB}\ee
which is same $T^2$ temperature scaling as the (1+1) Stefan-Boltzmann law \cite{Landsberg1989TheSC} describing black hole radiance \cite{Bekenstein:2003dt} and electron radiance \cite{Good:2022eub} as derived directly from the dynamics of the trajectory using the proper acceleration via Larmor power. 
 
While the quadratic scaling of temperature in Eq.~(\ref{SB}) describes thermal noise power transfer in one-dimensional optical systems \cite{Landsberg1989TheSC},
\be P = \frac{\pi}{6} \frac{k_B^2}{\hbar} T^2,\ee
the most widely known case of one-dimensional thermal radiation is Johnson noise or Nyquist noise of electrical circuits \cite{Nyquist:1928zz},
\be  P = \frac{\pi}{12} \frac{k_B^2}{\hbar} T^2,\ee
which is also proportional to temperature squared, yet with an emissivity of $\epsilon = 1/2$. This lower emissivity arises from the fact that photons in electrical networks are polarised, and thus the resistors act as grey bodies rather than black bodies.

\subsection{Stefan-Boltzmann from Entropy}
Consider the classical accelerating electron in thermal equilibrium with its environment. By the second law, the probability distribution, $p(n)$, must be such as to maximize the system entropy. Following Oliver \cite{Nyquist:1928zz}, we will determine $p(n)$ where $n$ is an integer.  Here $n$ is the number of photons emitted by the ensemble system. We start with the definition of Gibbs entropy for the electron,
\be S_e = -k_B \sum_{n=0}^{\infty} p(n) \ln p(n).\ee
with constraints of unitarity and averaging
\be \sum_{n=0}^{\infty} p(n) = 1, \qquad \sum_{n=0}^{\infty} n\; p(n) = \bar{n}.\ee
The first constraint demands $n$ must be some integer. The second constraint gives the average number of photons present where $\bar{n}$ need not be an integer.

The above summations will not vary as the distribution is varied as long as the entropy is maximized. A linear sum of all three
\be \sum p(n) \ln p(n) + \alpha \sum n \; p(n) + \beta \sum p(n), \ee
where $\alpha$ and $\beta$ are constants, will have zero variation
\be \sum \left[\ln p(n) + 1 + \alpha n + \beta \right] \delta p(n) = 0,\ee
for small perturbations $\delta p(n)$ of $p(n)$.  This is satisfied if
\be \ln p(n) + 1 + \alpha n + \beta = 0,\ee
which gives a probability distribution,
\be p(n) = e^{-1 - \beta} e^{-\alpha n}.\ee
Using the averaging and unitarity constrain, we solve
\be p(n) = \left(1-e^{-\alpha }\right) e^{-n \alpha }, \qquad \bar{n} = \frac{1}{e^{\alpha }-1}.\ee
Using the distribution $p(n)$ in the entropy, we find
\be \frac{S_e}{k_B} = \frac{\ln \left(e^{\alpha }-1\right)}{e^{\alpha }-1} - \frac{\ln \left(-e^{-\alpha }+1\right)}{-e^{-\alpha }+1}. \ee
The electron can be imagined to absorb an average non-integer classical energy (no $\hbar$)  
\be \bar{W}(\omega) = \bar{n} \mu_0 c e^2 \omega,\ee
in analog to $\bar{n}\hbar \omega$, suitable%
\footnote{For a gray body with absorptivity $a$, average absorbed radiation is $\bar{n}\hbar \omega \cdot a$. In our setup $a \sim \mu_0 c e^2 / \hbar$, see Sec.~\ref{sec:gray}.}
for the Stoney scale (nominally $\mu_0 c e^2$ is as we have seen, more than ten times smaller than $\hbar$).  This average energy comes from the surrounding outside thermal environment, producing an entropy change:
\be S_o = - \frac{\bar{n}\mu_0 c e^2\omega }{T} = k_B \frac{1}{e^{\alpha }-1} M,\ee
in the rest of the system `outside' the electron. The dimensionless $M$ is the Stoney temperature scale version of $\hbar \omega/(k_B T)$:
\be M \equiv \frac{\mu_0 c e^2}{k_B T} \omega. \ee 
The total change in entropy, $\Delta S = S_e + S_o$ will progress until, in equilibrium  $\Delta S$ is maximized. 
Taking a derivative of $\Delta S$ with respect to $\alpha$, gives 
\be \frac{1}{4} \text{csch}^2\left(\frac{\alpha }{2}\right) (M-\alpha ) = 0.\ee
This is true if $\alpha = M$.  The probability distribution and $\bar{n}$ is then written,
\be p(n) = \left(1-e^{-M}\right)e^{-n M }, \qquad \bar{n} = \frac{1}{e^{M }-1}.\ee
 Using $\bar{n}$, as found above, with $\alpha = M$, the average classical energy absorbed by the electron is
\be \bar{W}(\omega) = \bar{n} \mu c e^2 \omega = \frac{1}{e^M - 1} M k_B T.\ee
The total thermal power is 
\be P = \int_0^\infty \bar{W}(\omega) \frac{\diff \omega}{2\pi} = \frac{(k_B T)^2}{\mu_0 c e^2}\frac{1}{2\pi} \int_0^\infty \frac{M}{e^M-1} \diff M.\ee 
whose integral is $\pi^2/6$, so that
\be P = \frac{\pi}{12} \frac{k_B^2}{\mu_0 c e^2} T^2,\ee
which scales as the (1+1)D Stefan-Boltzmann law. This is classical thermal noise from a single accelerating electron \cite{Ievlev:2023inj}.

\subsection{Stefan-Boltzmann from Thermodynamics}

The Stefan-Boltzmann law can be derived from pure thermodynamics in two steps (see the original paper \cite{Stefan:1879txg}).
In this derivation we will not assume any particular form of the spectral frequency distribution.

\paragraph{Maxwell relations}

First, consider the Maxwell relations for the entropy. Let $U$ be the radiation energy, then $U = u(T) \, V$, where $u(T)$ is the energy density (we suppose that is depends only on the temperature $T$). Then we have:
\begin{equation}
	dU = T \, dS - p \, dV \,,
\end{equation}
from which it follows that
\begin{equation}
	dS = \frac{1}{T} (dU + p \, dV) = \frac{1}{T} \left( V \dv{u}{T} dT + (u + p) dV \right) \,.
\end{equation}
From this we can read off the first derivatives
\begin{equation}
\begin{aligned}
	\left( \pdv{S}{T} \right)_V &= \frac{V}{T} \dv{u}{T} \,, \\
	\left( \pdv{S}{V} \right)_T &= \frac{u + p}{T} \,.
\end{aligned}
\end{equation}
Computing the second derivative $\pdv*{S}{T}{V}$ in two different ways we obtain after some algebra
\begin{equation}
	\left( \pdv{p}{T} \right)_V = \frac{u + p}{T} \,.
\label{maxwell_result}
\end{equation}
To finish the derivation we need an equation of state. Let us derive it.

\paragraph{Equation of state in 3+1}

\begin{figure}
	\centering
	\includegraphics[width=0.4\linewidth]{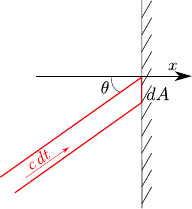}
	\caption{
		Radiation in a box
	}
	\label{fig:rad}
\end{figure}

We start with the 3+1 dimensional case.

Consider the radiation inside a perfectly reflecting box\footnote{The perfect reflectivity assumption in actually not mandatory, but it simplifies the derivations.}, see Fig.~\ref{fig:rad}.
When radiation waves hits the box wall, it gets reflected and therefore transfers some of its momentum to the wall.
Take some area element $dA$ of the wall, and let the $x$ axis be perpendicular to the wall in this vicinity. 
Let $\theta$ be the angle between the $x$ axis and the wavevector of an incoming electromagnetic wave.
Total momentum of the radiation coming from the solid angle $d \Omega$ is given by the energy divided by $c$:
\begin{equation}
	   |\vec{\mathtt{P}}| 
		= \frac{1}{c} \cdot u \cdot c \, dt \, \cos\theta \cdot dA \cdot \frac{d \Omega}{4\pi}
		= u \cos\theta \frac{d \Omega}{4\pi} \, dA \, dt,
\end{equation}
(recall that the energy can be computed as the energy density times the volume).
The momentum transfer is twice the $x$-projection:
\begin{equation}
	\Delta \mathtt{P}_x = 2 |\vec{\mathtt{P}}| \cos\theta = 2 u \cos^2\theta \frac{d \Omega}{4\pi} \, dA \, dt \,.
\end{equation}
Dividing the momentum by $dt$ we get the force, and then dividing by $dA$ we obtain the pressure.
Integration over the solid angle one side of the wall finally gives
\begin{equation}
	p = \frac{2 u}{4\pi} \int\limits_0^{2\pi} d\phi \int\limits_0^{1} \cos^2\theta \, d \cos\theta 
		= \frac{1}{3} u \,.
\end{equation}
Plugging this into \eqref{maxwell_result} yields a simple ODE whose solution
\begin{equation}
	u = C \cdot T^4,
\end{equation}
is determined only up to an arbitrary constant $C$. Note that on general grounds this constant must be positive, but otherwise it is not restricted by this derivation.

\paragraph{Other dimensions}

What happens in lower dimensions?

Consider a lower dimensional system embedded in the 3d space. This means that we can still consider electromagnetic waves even for the 1d system.
Then the only thing that we should modify in our derivation is the solid angle part. 

When the space is two-dimensional, there is no solid angle, and $\theta$ represents the polar angle. The integration in this case gives the result $p = u/2$.
In a one-dimensional system there are no angles at all, and we simply get $p = u$.

From this we can derive the corresponding Stefan-Boltzmann laws. We have summarized the results in Table~\ref{table:SB}.
For a body immersed in an equilibrium radiation heat bath the radiated energy per second is proportional to the energy density of the ambient radiation (the proportionality coefficient depends on the surface area and absorptivity).

\begin{table}[h]
\centering
\begin{tabular}{c | c | c}
	dim & eqn of state & Stefan-Boltzmann law \\
	\hline
	1 & $p = u$ & $P \sim T^2$ \\[1ex]
	2 & $p = \frac{1}{2} u$ & $P  \sim T^3$ \\[1ex]
	3 & $p = \frac{1}{3} u$ & $P  \sim T^4$ 
\end{tabular}
\caption{Stefan-Boltzmann law in various dimensions}
\label{table:SB}
\end{table}

\paragraph{Lessons}

What we just found is that from pure thermodynamics the Stephan-Boltzmann law is determined by the dimensionality of the space, up to a coefficient.
The coefficient is undetermined in this derivation; it depends on the  physical system under consideration and is not fixed on general grounds.
In particular, Eq.~\eqref{SB} is perfectly acceptable because of the scaling.

It is pleasing to find that Eq.~\eqref{SB} represents the Stephan-Boltzmann law in one-dimensional space, exactly what one would expect from a moving mirror in one spatial and one temporal dimension.

\subsection{Electron as a gray body}
\label{sec:gray}

Usually the Stephan-Boltzmann law relates the total power to the temperature. In 1+1 dimensions \cite{Landsberg1989TheSC},
%
\begin{equation}
    P_\text{black body} = \frac{\pi k_B^2}{6 \hbar } A T^2 \,,
\label{SB_bb}   
\end{equation}
see e.g. below Eq.~(10) in \cite{Landsberg1989TheSC}. 
Here, $A$ is the body's \textquote{surface area}: in 1+1 we have $A=1$ (one side) or $A=2$ (two sides).
In the moving mirror setup it is natural to take $A=1$, as the observer is usually on one (the right) side of the mirror.

We can compare Eq.~\eqref{SB_bb} to the power in our setup.
The quantum radiation power in Eq.~\eqref{SIPq},
\begin{equation}
    P_q = \frac{\pi k_B^2}{12\hbar}T^2 \,,
\end{equation}
we see that it corresponds to a gray body with absorptivity $a = 1/2$.

The classical radiation power in Eq.~\eqref{SBlaw}, when re-expressed in terms of the quantum temperature scale (cf. Eq.~\eqref{bridge}) becomes
\begin{equation}
	\bar{P}_e =  \frac{\mu_0  e^2 \kappa^2}{48 \pi c} 
		= \frac{\mu_0 c e^2}{\hbar} \frac{\pi}{12} \frac{k_B^2}{\hbar} T^2 \,,
\label{SB_el}
\end{equation}
which is also proportional to the square of the temperature $T^2$.
We see that Eq.~\eqref{SB_bb} and Eq.~\eqref{SB_el} are not quite the same, they are off by a factor (we take $A=1$, which is natural from the mirror's perspective):
\begin{equation}
	\bar{P}_e = \frac{\mu_0 c e^2}{2 \hbar} P_\text{black body} \,, \quad
	\frac{\mu_0 c e^2}{2 \hbar} \approx 0.0458 \,.
\end{equation}
The fact that there is a coefficient tells us that the electron looks like a 1+1 quantum-radiating gray body that absorbs only $\sim 4.58\%$ of the incoming radiation.

The physical meaning of this absorptivity is intriguing and deserves further investigation, which we leave for a future work.

\section{Conclusion}
\label{sec:concl}
In this note, we have helped develop the analogy between the dynamical Casimir, black hole, and electron radiation temperature. We have found periods of constant power and radiation reaction, indicative of thermal equilibrium.  Indeed, by analogy with the dynamical Casimir effect, we have demonstrated a useful notion of thermality by the symmetry between frequency modes in the analog spectrum for the radiation of an accelerated electron, which at ultra-relativistic speeds manifests an explicit uniform radiation emission commensurate with the spectral Planck distribution.  The constant temperature is intuitively consistent with the constant periods of power, self-force, and peel acceleration.
\acknowledgments{Thanks is given to Paul Davies, Eric Linder, and Morgan Lynch for insightful discussion. Funding comes in part from the FY2021-SGP-1-STMM Faculty Development Competitive Research Grant No. 021220FD3951 at Nazarbayev University.} 

\bibliography{main} 

\begin{thebibliography}{49}%
\makeatletter
\providecommand \@ifxundefined [1]{%
 \@ifx{#1\undefined}
}%
\providecommand \@ifnum [1]{%
 \ifnum #1\expandafter \@firstoftwo
 \else \expandafter \@secondoftwo
 \fi
}%
\providecommand \@ifx [1]{%
 \ifx #1\expandafter \@firstoftwo
 \else \expandafter \@secondoftwo
 \fi
}%
\providecommand \natexlab [1]{#1}%
\providecommand \enquote  [1]{``#1''}%
\providecommand \bibnamefont  [1]{#1}%
\providecommand \bibfnamefont [1]{#1}%
\providecommand \citenamefont [1]{#1}%
\providecommand \href@noop [0]{\@secondoftwo}%
\providecommand \href [0]{\begingroup \@sanitize@url \@href}%
\providecommand \@href[1]{\@@startlink{#1}\@@href}%
\providecommand \@@href[1]{\endgroup#1\@@endlink}%
\providecommand \@sanitize@url [0]{\catcode `\\12\catcode `\$12\catcode
  `\&12\catcode `\#12\catcode `\^12\catcode `\_12\catcode `\%12\relax}%
\providecommand \@@startlink[1]{}%
\providecommand \@@endlink[0]{}%
\providecommand \url  [0]{\begingroup\@sanitize@url \@url }%
\providecommand \@url [1]{\endgroup\@href {#1}{\urlprefix }}%
\providecommand \urlprefix  [0]{URL }%
\providecommand \Eprint [0]{\href }%
\providecommand \doibase [0]{http://dx.doi.org/}%
\providecommand \selectlanguage [0]{\@gobble}%
\providecommand \bibinfo  [0]{\@secondoftwo}%
\providecommand \bibfield  [0]{\@secondoftwo}%
\providecommand \translation [1]{[#1]}%
\providecommand \BibitemOpen [0]{}%
\providecommand \bibitemStop [0]{}%
\providecommand \bibitemNoStop [0]{.\EOS\space}%
\providecommand \EOS [0]{\spacefactor3000\relax}%
\providecommand \BibitemShut  [1]{\csname bibitem#1\endcsname}%
\let\auto@bib@innerbib\@empty
\bibitem [{\citenamefont {Hawking}(1975)}]{Hawking:1974sw}%
  \BibitemOpen
  \bibfield  {author} {\bibinfo {author} {\bibfnamefont {S.W.}\ \bibnamefont
  {Hawking}},\ }\bibfield  {title} {\enquote {\bibinfo {title} {{Particle
  Creation by Black Holes}},}\ }\href {\doibase 10.1007/BF02345020} {\bibfield
  {journal} {\bibinfo  {journal} {Commun. Math. Phys.}\ }\textbf {\bibinfo
  {volume} {43}},\ \bibinfo {pages} {199--220} (\bibinfo {year}
  {1975})}\BibitemShut {NoStop}%
\bibitem [{\citenamefont {Bekenstein}\ and\ \citenamefont
  {Mayo}(2001)}]{Bekenstein:2001tj}%
  \BibitemOpen
  \bibfield  {author} {\bibinfo {author} {\bibfnamefont {Jacob~D.}\
  \bibnamefont {Bekenstein}}\ and\ \bibinfo {author} {\bibfnamefont
  {Avraham~E.}\ \bibnamefont {Mayo}},\ }\bibfield  {title} {\enquote {\bibinfo
  {title} {{Black holes are one-dimensional}},}\ }\href {\doibase
  10.1023/A:1015278813573} {\bibfield  {journal} {\bibinfo  {journal} {Gen.
  Rel. Grav.}\ }\textbf {\bibinfo {volume} {33}},\ \bibinfo {pages}
  {2095--2099} (\bibinfo {year} {2001})},\ \Eprint
  {http://arxiv.org/abs/gr-qc/0105055} {arXiv:gr-qc/0105055} \BibitemShut
  {NoStop}%
\bibitem [{\citenamefont {Good}\ and\ \citenamefont
  {Davies}(2023)}]{Good:2022eub}%
  \BibitemOpen
  \bibfield  {author} {\bibinfo {author} {\bibfnamefont {Michael R.~R.}\
  \bibnamefont {Good}}\ and\ \bibinfo {author} {\bibfnamefont {Paul C.~W.}\
  \bibnamefont {Davies}},\ }\bibfield  {title} {\enquote {\bibinfo {title}
  {{Infrared acceleration radiation}},}\ }\href {\doibase
  10.1007/s10701-023-00694-x} {\bibfield  {journal} {\bibinfo  {journal}
  {Foundations of Physics}\ }\textbf {\bibinfo {volume} {53}},\ \bibinfo
  {pages} {53} (\bibinfo {year} {2023})},\ \Eprint
  {http://arxiv.org/abs/2206.07291} {arXiv:2206.07291 [gr-qc]} \BibitemShut
  {NoStop}%
\bibitem [{\citenamefont {Stoney}(1881)}]{stoney1881physical}%
  \BibitemOpen
  \bibfield  {author} {\bibinfo {author} {\bibfnamefont {George~Johnstone}\
  \bibnamefont {Stoney}},\ }\bibfield  {title} {\enquote {\bibinfo {title} {On
  the physical units of nature},}\ }\href@noop {} {\bibfield  {journal}
  {\bibinfo  {journal} {Scientific Proceedings of the Royal Dublin Society}\
  }\textbf {\bibinfo {volume} {3}},\ \bibinfo {pages} {51--60} (\bibinfo {year}
  {1881})}\BibitemShut {NoStop}%
\bibitem [{\citenamefont {{Barrow}}(1983)}]{barrowSTONE}%
  \BibitemOpen
  \bibfield  {author} {\bibinfo {author} {\bibfnamefont {J.~D.}\ \bibnamefont
  {{Barrow}}},\ }\bibfield  {title} {\enquote {\bibinfo {title} {{Natural Units
  Before Planck}},}\ }\href
  {https://adsabs.harvard.edu/full/1983QJRAS..24...24B} {\bibfield  {journal}
  {\bibinfo  {journal} {Quarterly Journal of the Royal Astronomical Society}\
  }\textbf {\bibinfo {volume} {24}},\ \bibinfo {pages} {24--26} (\bibinfo
  {year} {1983})}\BibitemShut {NoStop}%
\bibitem [{\citenamefont {Ievlev}\ and\ \citenamefont
  {Good}(2023)}]{Ievlev:2023inj}%
  \BibitemOpen
  \bibfield  {author} {\bibinfo {author} {\bibfnamefont {E.}~\bibnamefont
  {Ievlev}}\ and\ \bibinfo {author} {\bibfnamefont {Michael R.~R.}\
  \bibnamefont {Good}},\ }\bibfield  {title} {\enquote {\bibinfo {title}
  {{Thermal Larmor radiation}},}\ }\href@noop {} {\  (\bibinfo {year}
  {2023})},\ \Eprint {http://arxiv.org/abs/2303.03676} {arXiv:2303.03676
  [gr-qc]} \BibitemShut {NoStop}%
\bibitem [{\citenamefont {Moore}(1970)}]{moore1970quantum}%
  \BibitemOpen
  \bibfield  {author} {\bibinfo {author} {\bibfnamefont {Gerald~T.}\
  \bibnamefont {Moore}},\ }\bibfield  {title} {\enquote {\bibinfo {title}
  {Quantum theory of the electromagnetic field in a variable‐length
  one‐dimensional cavity},}\ }\href {https://doi.org/10.1063/1.1665432}
  {\bibfield  {journal} {\bibinfo  {journal} {J. of Math. Phys.}\ }\textbf
  {\bibinfo {volume} {11}},\ \bibinfo {pages} {2679--2691} (\bibinfo {year}
  {1970})}\BibitemShut {NoStop}%
\bibitem [{\citenamefont {DeWitt}(1975)}]{DeWitt:1975ys}%
  \BibitemOpen
  \bibfield  {author} {\bibinfo {author} {\bibfnamefont {Bryce~S.}\
  \bibnamefont {DeWitt}},\ }\bibfield  {title} {\enquote {\bibinfo {title}
  {{Quantum Field Theory in Curved Space-Time}},}\ }\href {\doibase
  10.1016/0370-1573(75)90051-4} {\bibfield  {journal} {\bibinfo  {journal}
  {Phys. Rept.}\ }\textbf {\bibinfo {volume} {19}},\ \bibinfo {pages}
  {295--357} (\bibinfo {year} {1975})}\BibitemShut {NoStop}%
\bibitem [{\citenamefont {Davies}\ and\ \citenamefont
  {Fulling}(1977)}]{Davies:1977yv}%
  \BibitemOpen
  \bibfield  {author} {\bibinfo {author} {\bibfnamefont {P.C.W.}\ \bibnamefont
  {Davies}}\ and\ \bibinfo {author} {\bibfnamefont {S.A.}\ \bibnamefont
  {Fulling}},\ }\bibfield  {title} {\enquote {\bibinfo {title} {{Radiation from
  Moving Mirrors and from Black Holes}},}\ }\href {\doibase
  10.1098/rspa.1977.0130} {\bibfield  {journal} {\bibinfo  {journal} {Proc. R.
  Soc. Lond. A}\ }\textbf {\bibinfo {volume} {A356}},\ \bibinfo {pages}
  {237--257} (\bibinfo {year} {1977})}\BibitemShut {NoStop}%
\bibitem [{\citenamefont {Ritus}(2003)}]{Ritus:2003wu}%
  \BibitemOpen
  \bibfield  {author} {\bibinfo {author} {\bibfnamefont {V.I.}\ \bibnamefont
  {Ritus}},\ }\bibfield  {title} {\enquote {\bibinfo {title} {{The Symmetry,
  inferable from Bogoliubov transformation, between the processes induced by
  the mirror in two-dimensional and the charge in four-dimensional
  space-time}},}\ }\href {\doibase 10.1134/1.1600792} {\bibfield  {journal}
  {\bibinfo  {journal} {J. Exp. Theor. Phys.}\ }\textbf {\bibinfo {volume}
  {97}},\ \bibinfo {pages} {10--23} (\bibinfo {year} {2003})},\ \Eprint
  {http://arxiv.org/abs/hep-th/0309181} {arXiv:hep-th/0309181} \BibitemShut
  {NoStop}%
\bibitem [{\citenamefont {Ritus}(2002)}]{Ritus:2002rq}%
  \BibitemOpen
  \bibfield  {author} {\bibinfo {author} {\bibfnamefont {V.I.}\ \bibnamefont
  {Ritus}},\ }\bibfield  {title} {\enquote {\bibinfo {title} {{Vacuum-vacuum
  amplitudes in the theory of quantum radiation by mirrors in 1+1-space and
  charges in 3+1-space}},}\ }\href {\doibase 10.1142/S0217751X02010467}
  {\bibfield  {journal} {\bibinfo  {journal} {Int. J. Mod. Phys. A}\ }\textbf
  {\bibinfo {volume} {17}},\ \bibinfo {pages} {1033--1040} (\bibinfo {year}
  {2002})}\BibitemShut {NoStop}%
\bibitem [{\citenamefont {Ritus}(1998)}]{Ritus:1999eu}%
  \BibitemOpen
  \bibfield  {author} {\bibinfo {author} {\bibfnamefont {V.I.}\ \bibnamefont
  {Ritus}},\ }\bibfield  {title} {\enquote {\bibinfo {title} {{Symmetries and
  causes of the coincidence of the radiation spectra of mirrors and charges in
  (1+1) and (3+1) spaces}},}\ }\href {\doibase 10.1134/1.558646} {\bibfield
  {journal} {\bibinfo  {journal} {J. Exp. Theor. Phys.}\ }\textbf {\bibinfo
  {volume} {87}},\ \bibinfo {pages} {25--34} (\bibinfo {year} {1998})},\
  \Eprint {http://arxiv.org/abs/hep-th/9903083} {arXiv:hep-th/9903083}
  \BibitemShut {NoStop}%
\bibitem [{\citenamefont {Nikishov}\ and\ \citenamefont
  {Ritus}(1995)}]{Nikishov:1995qs}%
  \BibitemOpen
  \bibfield  {author} {\bibinfo {author} {\bibfnamefont {A.I.}\ \bibnamefont
  {Nikishov}}\ and\ \bibinfo {author} {\bibfnamefont {V.I.}\ \bibnamefont
  {Ritus}},\ }\bibfield  {title} {\enquote {\bibinfo {title} {{Emission of
  scalar photons by an accelerated mirror in (1+1) space and its relation to
  the radiation from an electrical charge in classical electrodynamics}},}\
  }\href@noop {} {\bibfield  {journal} {\bibinfo  {journal} {J. Exp. Theor.
  Phys.}\ }\textbf {\bibinfo {volume} {81}},\ \bibinfo {pages} {615--624}
  (\bibinfo {year} {1995})}\BibitemShut {NoStop}%
\bibitem [{\citenamefont {Zhakenuly}\ \emph {et~al.}(2021)\citenamefont
  {Zhakenuly}, \citenamefont {Temirkhan}, \citenamefont {Good},\ and\
  \citenamefont {Chen}}]{Zhakenuly:2021pfm}%
  \BibitemOpen
  \bibfield  {author} {\bibinfo {author} {\bibfnamefont {Abay}\ \bibnamefont
  {Zhakenuly}}, \bibinfo {author} {\bibfnamefont {Maksat}\ \bibnamefont
  {Temirkhan}}, \bibinfo {author} {\bibfnamefont {Michael R.~R.}\ \bibnamefont
  {Good}}, \ and\ \bibinfo {author} {\bibfnamefont {Pisin}\ \bibnamefont
  {Chen}},\ }\bibfield  {title} {\enquote {\bibinfo {title} {{Quantum power
  distribution of relativistic acceleration radiation: classical electrodynamic
  analogies with perfectly reflecting moving mirrors}},}\ }\href {\doibase
  10.3390/sym13040653} {\bibfield  {journal} {\bibinfo  {journal} {Symmetry}\
  }\textbf {\bibinfo {volume} {13}},\ \bibinfo {pages} {653} (\bibinfo {year}
  {2021})},\ \Eprint {http://arxiv.org/abs/2101.02511} {arXiv:2101.02511
  [gr-qc]} \BibitemShut {NoStop}%
\bibitem [{\citenamefont {Good}\ and\ \citenamefont
  {Linder}(2022)}]{Good:2021ffo}%
  \BibitemOpen
  \bibfield  {author} {\bibinfo {author} {\bibfnamefont {Michael R.~R.}\
  \bibnamefont {Good}}\ and\ \bibinfo {author} {\bibfnamefont {Eric~V.}\
  \bibnamefont {Linder}},\ }\bibfield  {title} {\enquote {\bibinfo {title}
  {{Quantum power: a Lorentz invariant approach to Hawking radiation}},}\
  }\href {\doibase 10.1140/epjc/s10052-022-10167-6} {\bibfield  {journal}
  {\bibinfo  {journal} {Eur. Phys. J. C}\ }\textbf {\bibinfo {volume} {82}},\
  \bibinfo {pages} {204} (\bibinfo {year} {2022})},\ \Eprint
  {http://arxiv.org/abs/2111.15148} {arXiv:2111.15148 [gr-qc]} \BibitemShut
  {NoStop}%
\bibitem [{\citenamefont {Ford}\ and\ \citenamefont
  {Vilenkin}(1982)}]{Ford:1982ct}%
  \BibitemOpen
  \bibfield  {author} {\bibinfo {author} {\bibfnamefont {L.H.}\ \bibnamefont
  {Ford}}\ and\ \bibinfo {author} {\bibfnamefont {Alexander}\ \bibnamefont
  {Vilenkin}},\ }\bibfield  {title} {\enquote {\bibinfo {title} {{Quantum
  radiation by moving mirrors}},}\ }\href {\doibase 10.1103/PhysRevD.25.2569}
  {\bibfield  {journal} {\bibinfo  {journal} {Phys. Rev. D}\ }\textbf {\bibinfo
  {volume} {25}},\ \bibinfo {pages} {2569} (\bibinfo {year}
  {1982})}\BibitemShut {NoStop}%
\bibitem [{\citenamefont {Unruh}\ and\ \citenamefont
  {Wald}(1982)}]{Unruh:1982ic}%
  \BibitemOpen
  \bibfield  {author} {\bibinfo {author} {\bibfnamefont {W.~G.}\ \bibnamefont
  {Unruh}}\ and\ \bibinfo {author} {\bibfnamefont {Robert~M.}\ \bibnamefont
  {Wald}},\ }\bibfield  {title} {\enquote {\bibinfo {title} {{Acceleration
  Radiation and Generalized Second Law of Thermodynamics}},}\ }\href {\doibase
  10.1103/PhysRevD.25.942} {\bibfield  {journal} {\bibinfo  {journal} {Phys.
  Rev. D}\ }\textbf {\bibinfo {volume} {25}},\ \bibinfo {pages} {942--958}
  (\bibinfo {year} {1982})}\BibitemShut {NoStop}%
\bibitem [{\citenamefont {Myrzakul}\ \emph {et~al.}(2021)\citenamefont
  {Myrzakul}, \citenamefont {Xiong},\ and\ \citenamefont
  {Good}}]{Myrzakul:2021bgj}%
  \BibitemOpen
  \bibfield  {author} {\bibinfo {author} {\bibfnamefont {Aizhan}\ \bibnamefont
  {Myrzakul}}, \bibinfo {author} {\bibfnamefont {Chi}\ \bibnamefont {Xiong}}, \
  and\ \bibinfo {author} {\bibfnamefont {Michael R.~R.}\ \bibnamefont {Good}},\
  }\bibfield  {title} {\enquote {\bibinfo {title} {{CGHS Black Hole Analog
  Moving Mirror and Its Relativistic Quantum Information as Radiation
  Reaction}},}\ }\href {\doibase 10.3390/e23121664} {\bibfield  {journal}
  {\bibinfo  {journal} {Entropy}\ }\textbf {\bibinfo {volume} {23}},\ \bibinfo
  {pages} {1664} (\bibinfo {year} {2021})},\ \Eprint
  {http://arxiv.org/abs/2101.08139} {arXiv:2101.08139 [gr-qc]} \BibitemShut
  {NoStop}%
\bibitem [{\citenamefont {Zangwill}(2013)}]{Zangwill:1507229}%
  \BibitemOpen
  \bibfield  {author} {\bibinfo {author} {\bibfnamefont {Andrew}\ \bibnamefont
  {Zangwill}},\ }\href {https://cds.cern.ch/record/1507229} {\emph {\bibinfo
  {title} {{Modern electrodynamics}}}}\ (\bibinfo  {publisher} {Cambridge Univ.
  Press},\ \bibinfo {address} {Cambridge},\ \bibinfo {year} {2013})\BibitemShut
  {NoStop}%
\bibitem [{\citenamefont {Jackson}(1999)}]{Jackson:490457}%
  \BibitemOpen
  \bibfield  {author} {\bibinfo {author} {\bibfnamefont {John~David}\
  \bibnamefont {Jackson}},\ }\href {https://cds.cern.ch/record/490457} {\emph
  {\bibinfo {title} {{Classical electrodynamics; 3rd ed.}}}}\ (\bibinfo
  {publisher} {Wiley},\ \bibinfo {address} {New York, NY},\ \bibinfo {year}
  {1999})\BibitemShut {NoStop}%
\bibitem [{\citenamefont {Chang}\ and\ \citenamefont
  {Falkoff}(1949)}]{PhysRev.76.365}%
  \BibitemOpen
  \bibfield  {author} {\bibinfo {author} {\bibfnamefont {C.~S.~Wang}\
  \bibnamefont {Chang}}\ and\ \bibinfo {author} {\bibfnamefont {D.~L.}\
  \bibnamefont {Falkoff}},\ }\bibfield  {title} {\enquote {\bibinfo {title} {On
  the continuous gamma-radiation accompanying the beta-decay of nuclei},}\
  }\href {\doibase 10.1103/PhysRev.76.365} {\bibfield  {journal} {\bibinfo
  {journal} {Phys. Rev.}\ }\textbf {\bibinfo {volume} {76}},\ \bibinfo {pages}
  {365--371} (\bibinfo {year} {1949})}\BibitemShut {NoStop}%
\bibitem [{\citenamefont {Ballagh}\ \emph {et~al.}(1983)\citenamefont {Ballagh}
  \emph {et~al.}}]{Ballagh:1983zr}%
  \BibitemOpen
  \bibfield  {author} {\bibinfo {author} {\bibfnamefont {H.~C.}\ \bibnamefont
  {Ballagh}} \emph {et~al.},\ }\bibfield  {title} {\enquote {\bibinfo {title}
  {{Observation of muon inner bremsstrahlung in deep inelastic neutrino
  scattering}},}\ }\href {\doibase 10.1103/PhysRevLett.50.1963} {\bibfield
  {journal} {\bibinfo  {journal} {Phys. Rev. Lett.}\ }\textbf {\bibinfo
  {volume} {50}},\ \bibinfo {pages} {1963--1966} (\bibinfo {year}
  {1983})}\BibitemShut {NoStop}%
\bibitem [{\citenamefont {Strominger}(2017)}]{Strominger:2017zoo}%
  \BibitemOpen
  \bibfield  {author} {\bibinfo {author} {\bibfnamefont {Andrew}\ \bibnamefont
  {Strominger}},\ }\bibfield  {title} {\enquote {\bibinfo {title} {{Lectures on
  the Infrared Structure of Gravity and Gauge Theory}},}\ }\href@noop {} {\
  (\bibinfo {year} {2017})},\ \Eprint {http://arxiv.org/abs/1703.05448}
  {arXiv:1703.05448 [hep-th]} \BibitemShut {NoStop}%
\bibitem [{\citenamefont {Cardoso}\ \emph {et~al.}(2003)\citenamefont
  {Cardoso}, \citenamefont {Lemos},\ and\ \citenamefont
  {Yoshida}}]{Cardoso:2003cn}%
  \BibitemOpen
  \bibfield  {author} {\bibinfo {author} {\bibfnamefont {Vitor}\ \bibnamefont
  {Cardoso}}, \bibinfo {author} {\bibfnamefont {Jose P.~S.}\ \bibnamefont
  {Lemos}}, \ and\ \bibinfo {author} {\bibfnamefont {Shijun}\ \bibnamefont
  {Yoshida}},\ }\bibfield  {title} {\enquote {\bibinfo {title}
  {{Electromagnetic radiation from collisions at almost the speed of light: An
  Extremely relativistic charged particle falling into a Schwarzschild black
  hole}},}\ }\href {\doibase 10.1103/PhysRevD.68.084011} {\bibfield  {journal}
  {\bibinfo  {journal} {Phys. Rev. D}\ }\textbf {\bibinfo {volume} {68}},\
  \bibinfo {pages} {084011} (\bibinfo {year} {2003})},\ \Eprint
  {http://arxiv.org/abs/gr-qc/0307104} {arXiv:gr-qc/0307104} \BibitemShut
  {NoStop}%
\bibitem [{\citenamefont {Tomaras}\ and\ \citenamefont
  {Toumbas}(2020)}]{Tomaras}%
  \BibitemOpen
  \bibfield  {author} {\bibinfo {author} {\bibfnamefont {Theodore~N.}\
  \bibnamefont {Tomaras}}\ and\ \bibinfo {author} {\bibfnamefont {Nicolaos}\
  \bibnamefont {Toumbas}},\ }\bibfield  {title} {\enquote {\bibinfo {title} {Ir
  dynamics and entanglement entropy},}\ }\href {\doibase
  10.1103/PhysRevD.101.065006} {\bibfield  {journal} {\bibinfo  {journal}
  {Phys. Rev. D}\ }\textbf {\bibinfo {volume} {101}},\ \bibinfo {pages}
  {065006} (\bibinfo {year} {2020})}\BibitemShut {NoStop}%
\bibitem [{\citenamefont {Good}(2017)}]{Good:2016yht}%
  \BibitemOpen
  \bibfield  {author} {\bibinfo {author} {\bibfnamefont {Michael R.~R.}\
  \bibnamefont {Good}},\ }\href {https://arxiv.org/abs/1612.02459} {\emph
  {\bibinfo {title} {{Reflecting at the Speed of Light}}}}\ (\bibinfo
  {publisher} {World Scientific},\ \bibinfo {address} {Singapore},\ \bibinfo
  {year} {2017})\BibitemShut {NoStop}%
\bibitem [{\citenamefont {Wilczek}(1993)}]{wilczek1993quantum}%
  \BibitemOpen
  \bibfield  {author} {\bibinfo {author} {\bibfnamefont {Frank}\ \bibnamefont
  {Wilczek}},\ }\bibfield  {title} {\enquote {\bibinfo {title} {{Quantum purity
  at a small price: Easing a black hole paradox}},}\ }in\ \href@noop {} {\emph
  {\bibinfo {booktitle} {{International Symposium on Black holes, Membranes,
  Wormholes and Superstrings}}}}\ (\bibinfo {year} {1993})\ pp.\ \bibinfo
  {pages} {1--21},\ \Eprint {http://arxiv.org/abs/hep-th/9302096}
  {arXiv:hep-th/9302096} \BibitemShut {NoStop}%
\bibitem [{\citenamefont {Landsberg}\ and\ \citenamefont
  {Vos}(1989)}]{Landsberg1989TheSC}%
  \BibitemOpen
  \bibfield  {author} {\bibinfo {author} {\bibfnamefont {P~T}\ \bibnamefont
  {Landsberg}}\ and\ \bibinfo {author} {\bibfnamefont {A~De}\ \bibnamefont
  {Vos}},\ }\bibfield  {title} {\enquote {\bibinfo {title} {The
  stefan-boltzmann constant in n-dimensional space},}\ }\href {\doibase
  10.1088/0305-4470/22/8/021} {\bibfield  {journal} {\bibinfo  {journal}
  {Journal of Physics A: Mathematical and General}\ }\textbf {\bibinfo {volume}
  {22}},\ \bibinfo {pages} {1073} (\bibinfo {year} {1989})}\BibitemShut
  {NoStop}%
\bibitem [{\citenamefont {Feynman}(1996)}]{Feynman:1996kb}%
  \BibitemOpen
  \bibfield  {author} {\bibinfo {author} {\bibfnamefont {R.~P.}\ \bibnamefont
  {Feynman}},\ }\href@noop {} {\emph {\bibinfo {title} {{Feynman lectures on
  gravitation}}}},\ edited by\ \bibinfo {editor} {\bibfnamefont {F.~B.}\
  \bibnamefont {Morinigo}}, \bibinfo {editor} {\bibfnamefont {W.~G.}\
  \bibnamefont {Wagner}}, \ and\ \bibinfo {editor} {\bibfnamefont
  {B.}~\bibnamefont {Hatfield}}\ (\bibinfo {year} {1996})\BibitemShut {NoStop}%
\bibitem [{\citenamefont {Bianchi}\ and\ \citenamefont
  {Smerlak}(2014)}]{Bianchi:2014qua}%
  \BibitemOpen
  \bibfield  {author} {\bibinfo {author} {\bibfnamefont {Eugenio}\ \bibnamefont
  {Bianchi}}\ and\ \bibinfo {author} {\bibfnamefont {Matteo}\ \bibnamefont
  {Smerlak}},\ }\bibfield  {title} {\enquote {\bibinfo {title} {{Entanglement
  entropy and negative energy in two dimensions}},}\ }\href {\doibase
  10.1103/PhysRevD.90.041904} {\bibfield  {journal} {\bibinfo  {journal} {Phys.
  Rev. D}\ }\textbf {\bibinfo {volume} {90}},\ \bibinfo {pages} {041904}
  (\bibinfo {year} {2014})},\ \Eprint {http://arxiv.org/abs/1404.0602}
  {arXiv:1404.0602 [gr-qc]} \BibitemShut {NoStop}%
\bibitem [{\citenamefont {Barcelo}\ \emph {et~al.}(2011)\citenamefont
  {Barcelo}, \citenamefont {Liberati}, \citenamefont {Sonego},\ and\
  \citenamefont {Visser}}]{Barcelo:2010pj}%
  \BibitemOpen
  \bibfield  {author} {\bibinfo {author} {\bibfnamefont {Carlos}\ \bibnamefont
  {Barcelo}}, \bibinfo {author} {\bibfnamefont {Stefano}\ \bibnamefont
  {Liberati}}, \bibinfo {author} {\bibfnamefont {Sebastiano}\ \bibnamefont
  {Sonego}}, \ and\ \bibinfo {author} {\bibfnamefont {Matt}\ \bibnamefont
  {Visser}},\ }\bibfield  {title} {\enquote {\bibinfo {title} {{Minimal
  conditions for the existence of a Hawking-like flux}},}\ }\href {\doibase
  10.1103/PhysRevD.83.041501} {\bibfield  {journal} {\bibinfo  {journal} {Phys.
  Rev. D}\ }\textbf {\bibinfo {volume} {83}},\ \bibinfo {pages} {041501}
  (\bibinfo {year} {2011})},\ \Eprint {http://arxiv.org/abs/1011.5593}
  {arXiv:1011.5593 [gr-qc]} \BibitemShut {NoStop}%
\bibitem [{\citenamefont {Carlitz}\ and\ \citenamefont
  {Willey}(1987{\natexlab{a}})}]{CW2lifetime}%
  \BibitemOpen
  \bibfield  {author} {\bibinfo {author} {\bibfnamefont {Robert~D.}\
  \bibnamefont {Carlitz}}\ and\ \bibinfo {author} {\bibfnamefont {Raymond~S.}\
  \bibnamefont {Willey}},\ }\bibfield  {title} {\enquote {\bibinfo {title}
  {Lifetime of a black hole},}\ }\href {\doibase 10.1103/PhysRevD.36.2336}
  {\bibfield  {journal} {\bibinfo  {journal} {Phys. Rev. D}\ }\textbf {\bibinfo
  {volume} {36}},\ \bibinfo {pages} {2336--2341} (\bibinfo {year}
  {1987}{\natexlab{a}})}\BibitemShut {NoStop}%
\bibitem [{\citenamefont {Fulling}\ and\ \citenamefont
  {Davies}(1976)}]{Davies:1976hi}%
  \BibitemOpen
  \bibfield  {author} {\bibinfo {author} {\bibfnamefont {S.~A.}\ \bibnamefont
  {Fulling}}\ and\ \bibinfo {author} {\bibfnamefont {P.~C.~W.}\ \bibnamefont
  {Davies}},\ }\bibfield  {title} {\enquote {\bibinfo {title} {Radiation from a
  moving mirror in two dimensional space-time: conformal anomaly},}\ }\href
  {https://royalsocietypublishing.org/doi/abs/10.1098/rspa.1976.0045}
  {\bibfield  {journal} {\bibinfo  {journal} {Proc. R. Soc. Lond. A}\ }\textbf
  {\bibinfo {volume} {348}},\ \bibinfo {pages} {393--414} (\bibinfo {year}
  {1976})}\BibitemShut {NoStop}%
\bibitem [{\citenamefont {Fulling}(2005)}]{Fulling_optics}%
  \BibitemOpen
  \bibfield  {author} {\bibinfo {author} {\bibfnamefont {S.~A.}\ \bibnamefont
  {Fulling}},\ }\bibfield  {title} {\enquote {\bibinfo {title} {Review of some
  recent work on acceleration radiation},}\ }\href {\doibase
  10.1080/09500340500303637} {\bibfield  {journal} {\bibinfo  {journal} {J. of
  Mod. Optics}\ }\textbf {\bibinfo {volume} {52}},\ \bibinfo {pages}
  {2207--2213} (\bibinfo {year} {2005})}\BibitemShut {NoStop}%
\bibitem [{\citenamefont {Ritus}(2022)}]{Ritus:2022bph}%
  \BibitemOpen
  \bibfield  {author} {\bibinfo {author} {\bibfnamefont {V.~I.}\ \bibnamefont
  {Ritus}},\ }\bibfield  {title} {\enquote {\bibinfo {title} {{Finite value of
  the bare charge and the relation of the fine structure constant ratio for
  physical and bare charges to zero-point oscillations of the electromagnetic
  field in the vacuum}},}\ }\href {\doibase 10.3367/UFNe.2022.02.039167}
  {\bibfield  {journal} {\bibinfo  {journal} {Usp. Fiz. Nauk}\ }\textbf
  {\bibinfo {volume} {192}},\ \bibinfo {pages} {507--526} (\bibinfo {year}
  {2022})}\BibitemShut {NoStop}%
\bibitem [{\citenamefont {Ievlev}\ \emph {et~al.}(2023)\citenamefont {Ievlev},
  \citenamefont {Good},\ and\ \citenamefont {Linder}}]{Ievlev:2023bzk}%
  \BibitemOpen
  \bibfield  {author} {\bibinfo {author} {\bibfnamefont {Evgenii}\ \bibnamefont
  {Ievlev}}, \bibinfo {author} {\bibfnamefont {Michael R.~R.}\ \bibnamefont
  {Good}}, \ and\ \bibinfo {author} {\bibfnamefont {Eric~V.}\ \bibnamefont
  {Linder}},\ }\bibfield  {title} {\enquote {\bibinfo {title} {{Thermal
  Radiation from an Electron with Schwarzschild-Planck Acceleration}},}\
  }\href@noop {} {\  (\bibinfo {year} {2023})},\ \Eprint
  {http://arxiv.org/abs/2304.04412} {arXiv:2304.04412 [gr-qc]} \BibitemShut
  {NoStop}%
\bibitem [{\citenamefont {Cong}\ \emph {et~al.}(2019)\citenamefont {Cong},
  \citenamefont {Tjoa},\ and\ \citenamefont {Mann}}]{Cong:2018vqx}%
  \BibitemOpen
  \bibfield  {author} {\bibinfo {author} {\bibfnamefont {Wan}\ \bibnamefont
  {Cong}}, \bibinfo {author} {\bibfnamefont {Erickson}\ \bibnamefont {Tjoa}}, \
  and\ \bibinfo {author} {\bibfnamefont {Robert~B.}\ \bibnamefont {Mann}},\
  }\bibfield  {title} {\enquote {\bibinfo {title} {{Entanglement Harvesting
  with Moving Mirrors}},}\ }\href {\doibase 10.1007/JHEP06(2019)021} {\bibfield
   {journal} {\bibinfo  {journal} {JHEP}\ }\textbf {\bibinfo {volume} {06}},\
  \bibinfo {pages} {021} (\bibinfo {year} {2019})},\ \Eprint
  {http://arxiv.org/abs/1810.07359} {arXiv:1810.07359 [quant-ph]} \BibitemShut
  {NoStop}%
\bibitem [{\citenamefont {Good}\ \emph {et~al.}(2016)\citenamefont {Good},
  \citenamefont {Anderson},\ and\ \citenamefont {Evans}}]{Good:2016oey}%
  \BibitemOpen
  \bibfield  {author} {\bibinfo {author} {\bibfnamefont {Michael R.~R.}\
  \bibnamefont {Good}}, \bibinfo {author} {\bibfnamefont {Paul~R.}\
  \bibnamefont {Anderson}}, \ and\ \bibinfo {author} {\bibfnamefont
  {Charles~R.}\ \bibnamefont {Evans}},\ }\bibfield  {title} {\enquote {\bibinfo
  {title} {{Mirror Reflections of a Black Hole}},}\ }\href {\doibase
  10.1103/PhysRevD.94.065010} {\bibfield  {journal} {\bibinfo  {journal} {Phys.
  Rev. D}\ }\textbf {\bibinfo {volume} {94}},\ \bibinfo {pages} {065010}
  (\bibinfo {year} {2016})},\ \Eprint {http://arxiv.org/abs/1605.06635}
  {arXiv:1605.06635 [gr-qc]} \BibitemShut {NoStop}%
\bibitem [{\citenamefont {Good}(2018)}]{Good:2018zmx}%
  \BibitemOpen
  \bibfield  {author} {\bibinfo {author} {\bibfnamefont {Michael~R.R.}\
  \bibnamefont {Good}},\ }\bibfield  {title} {\enquote {\bibinfo {title}
  {{Spacetime Continuity and Quantum Information Loss}},}\ }\href {\doibase
  10.3390/universe4110122} {\bibfield  {journal} {\bibinfo  {journal}
  {Universe}\ }\textbf {\bibinfo {volume} {4}},\ \bibinfo {pages} {122}
  (\bibinfo {year} {2018})}\BibitemShut {NoStop}%
\bibitem [{\citenamefont {Good}\ and\ \citenamefont
  {Ong}(2020)}]{good2020particle}%
  \BibitemOpen
  \bibfield  {author} {\bibinfo {author} {\bibfnamefont {Michael~R.R.}\
  \bibnamefont {Good}}\ and\ \bibinfo {author} {\bibfnamefont {Yen~Chin}\
  \bibnamefont {Ong}},\ }\bibfield  {title} {\enquote {\bibinfo {title}
  {{Particle spectrum of the Reissner\textendash{}Nordstr\"om black hole}},}\
  }\href {\doibase 10.1140/epjc/s10052-020-08761-7} {\bibfield  {journal}
  {\bibinfo  {journal} {Eur. Phys. J. C}\ }\textbf {\bibinfo {volume} {80}},\
  \bibinfo {pages} {1169} (\bibinfo {year} {2020})},\ \Eprint
  {http://arxiv.org/abs/2004.03916} {arXiv:2004.03916 [gr-qc]} \BibitemShut
  {NoStop}%
\bibitem [{\citenamefont {Good}\ \emph {et~al.}(2021)\citenamefont {Good},
  \citenamefont {Foo},\ and\ \citenamefont {Linder}}]{Good:2020fjz}%
  \BibitemOpen
  \bibfield  {author} {\bibinfo {author} {\bibfnamefont {Michael R.~R.}\
  \bibnamefont {Good}}, \bibinfo {author} {\bibfnamefont {Joshua}\ \bibnamefont
  {Foo}}, \ and\ \bibinfo {author} {\bibfnamefont {Eric~V.}\ \bibnamefont
  {Linder}},\ }\bibfield  {title} {\enquote {\bibinfo {title} {{Accelerating
  boundary analog of a Kerr black hole}},}\ }\href {\doibase
  10.1088/1361-6382/abebb6} {\bibfield  {journal} {\bibinfo  {journal} {Class.
  Quant. Grav.}\ }\textbf {\bibinfo {volume} {38}},\ \bibinfo {pages} {085011}
  (\bibinfo {year} {2021})},\ \Eprint {http://arxiv.org/abs/2006.01349}
  {arXiv:2006.01349 [gr-qc]} \BibitemShut {NoStop}%
\bibitem [{\citenamefont {Carlitz}\ and\ \citenamefont
  {Willey}(1987{\natexlab{b}})}]{carlitz1987reflections}%
  \BibitemOpen
  \bibfield  {author} {\bibinfo {author} {\bibfnamefont {Robert~D.}\
  \bibnamefont {Carlitz}}\ and\ \bibinfo {author} {\bibfnamefont {Raymond~S.}\
  \bibnamefont {Willey}},\ }\bibfield  {title} {\enquote {\bibinfo {title}
  {Reflections on moving mirrors},}\ }\href {\doibase 10.1103/PhysRevD.36.2327}
  {\bibfield  {journal} {\bibinfo  {journal} {Phys. Rev. D}\ }\textbf {\bibinfo
  {volume} {36}},\ \bibinfo {pages} {2327--2335} (\bibinfo {year}
  {1987}{\natexlab{b}})}\BibitemShut {NoStop}%
\bibitem [{\citenamefont {Bell}\ and\ \citenamefont
  {Leinaas}(1983)}]{Bell:1982qr}%
  \BibitemOpen
  \bibfield  {author} {\bibinfo {author} {\bibfnamefont {J.~S.}\ \bibnamefont
  {Bell}}\ and\ \bibinfo {author} {\bibfnamefont {J.~M.}\ \bibnamefont
  {Leinaas}},\ }\bibfield  {title} {\enquote {\bibinfo {title} {{Electrons as
  accelerated thermometers}},}\ }\href {\doibase 10.1016/0550-3213(83)90601-6}
  {\bibfield  {journal} {\bibinfo  {journal} {Nucl. Phys. B}\ }\textbf
  {\bibinfo {volume} {212}},\ \bibinfo {pages} {131} (\bibinfo {year}
  {1983})}\BibitemShut {NoStop}%
\bibitem [{\citenamefont {Myhrvold}(1985)}]{Myhrvold:1983hv}%
  \BibitemOpen
  \bibfield  {author} {\bibinfo {author} {\bibfnamefont {Nathan~P.}\
  \bibnamefont {Myhrvold}},\ }\bibfield  {title} {\enquote {\bibinfo {title}
  {{Thermal Radiation From Accelerated Electrons}},}\ }\href {\doibase
  10.1016/0003-4916(85)90366-5} {\bibfield  {journal} {\bibinfo  {journal}
  {Annals Phys.}\ }\textbf {\bibinfo {volume} {160}},\ \bibinfo {pages} {102}
  (\bibinfo {year} {1985})}\BibitemShut {NoStop}%
\bibitem [{\citenamefont {Kolbenstvedt}(2001)}]{Kolbenstvedt_2001}%
  \BibitemOpen
  \bibfield  {author} {\bibinfo {author} {\bibfnamefont {H}~\bibnamefont
  {Kolbenstvedt}},\ }\bibfield  {title} {\enquote {\bibinfo {title}
  {Unruh-davies effect and the larmor radiation formula in hyperbolic
  motion},}\ }\href {\doibase 10.1238/Physica.Regular.063a00313} {\bibfield
  {journal} {\bibinfo  {journal} {Physica Scripta}\ }\textbf {\bibinfo {volume}
  {63}},\ \bibinfo {pages} {313} (\bibinfo {year} {2001})}\BibitemShut
  {NoStop}%
\bibitem [{\citenamefont {Landulfo}\ \emph {et~al.}(2019)\citenamefont
  {Landulfo}, \citenamefont {Fulling},\ and\ \citenamefont
  {Matsas}}]{Landulfo:2019tqj}%
  \BibitemOpen
  \bibfield  {author} {\bibinfo {author} {\bibfnamefont {Andr\'e~G.S.}\
  \bibnamefont {Landulfo}}, \bibinfo {author} {\bibfnamefont {Stephen~A.}\
  \bibnamefont {Fulling}}, \ and\ \bibinfo {author} {\bibfnamefont
  {George~E.A.}\ \bibnamefont {Matsas}},\ }\bibfield  {title} {\enquote
  {\bibinfo {title} {{Classical and quantum aspects of the radiation emitted by
  a uniformly accelerated charge: Larmor-Unruh reconciliation and
  zero-frequency Rindler modes}},}\ }\href {\doibase
  10.1103/PhysRevD.100.045020} {\bibfield  {journal} {\bibinfo  {journal}
  {Phys. Rev. D}\ }\textbf {\bibinfo {volume} {100}},\ \bibinfo {pages}
  {045020} (\bibinfo {year} {2019})},\ \Eprint
  {http://arxiv.org/abs/1907.06665} {arXiv:1907.06665 [gr-qc]} \BibitemShut
  {NoStop}%
\bibitem [{\citenamefont {Bekenstein}(2003)}]{Bekenstein:2003dt}%
  \BibitemOpen
  \bibfield  {author} {\bibinfo {author} {\bibfnamefont {Jacob~D.}\
  \bibnamefont {Bekenstein}},\ }\bibfield  {title} {\enquote {\bibinfo {title}
  {{Black holes and information theory}},}\ }\href {\doibase
  10.1080/00107510310001632523} {\bibfield  {journal} {\bibinfo  {journal}
  {Contemp. Phys.}\ }\textbf {\bibinfo {volume} {45}},\ \bibinfo {pages}
  {31--43} (\bibinfo {year} {2003})},\ \Eprint
  {http://arxiv.org/abs/quant-ph/0311049} {arXiv:quant-ph/0311049} \BibitemShut
  {NoStop}%
\bibitem [{\citenamefont {Nyquist}(1928)}]{Nyquist:1928zz}%
  \BibitemOpen
  \bibfield  {author} {\bibinfo {author} {\bibfnamefont {H.}~\bibnamefont
  {Nyquist}},\ }\bibfield  {title} {\enquote {\bibinfo {title} {{Thermal
  Agitation of Electric Charge in Conductors}},}\ }\href {\doibase
  10.1103/PhysRev.32.110} {\bibfield  {journal} {\bibinfo  {journal} {Phys.
  Rev.}\ }\textbf {\bibinfo {volume} {32}},\ \bibinfo {pages} {110--113}
  (\bibinfo {year} {1928})}\BibitemShut {NoStop}%
\bibitem [{\citenamefont {Stefan}(1879)}]{Stefan:1879txg}%
  \BibitemOpen
  \bibfield  {author} {\bibinfo {author} {\bibfnamefont {J.}~\bibnamefont
  {Stefan}},\ }\bibfield  {title} {\enquote {\bibinfo {title} {{\"Uber die
  Beziehung zwischen der W\"armestrahlung und der Temperatur}},}\ }\href@noop
  {} {\bibfield  {journal} {\bibinfo  {journal} {Sitzungsber. Kaiserl. Akad.
  Wiss. Math. Naturwiss. Cl. II. Abth.}\ }\textbf {\bibinfo {volume} {79}},\
  \bibinfo {pages} {391--428} (\bibinfo {year} {1879})}\BibitemShut {NoStop}%
\end{thebibliography}%
\end{document}